\documentclass[%
 reprint,
nofootinbib,
 amsmath,amssymb,
 aps,
prl,
longbibliography
]{revtex4-1}

\usepackage{graphicx}
\usepackage{dcolumn}
\usepackage{bm}
\usepackage{hyperref}
\usepackage{hhline,booktabs}

\begin{document}

\preprint{APS/123-QED}

\title{ Fermionic transport through a driven quantum point contact:  \\
breakdown of Floquet thermalization
beyond a critical driving frequency
}

\author{Ivan V. Dudinets}
\affiliation{Russian Quantum Center, 30 Bolshoy Boulevard, building 1, Skolkovo Innovation Center territory, Moscow, 121205, Russia}
\affiliation{Moscow Institute of Physics and Technology, Institutskii per. 9, Dolgoprudnyi, 141700, Russia}

\author{Oleg Lychkovskiy}
\affiliation{Skolkovo Institute of Science and Technology, the territory of the Skolkovo Innovation Center, Bolshoy Boulevard, 30, p.1, Moscow 121205, Russia}
\affiliation{Steklov Mathematical Institute of Russian Academy of Sciences, Gubkina str., 8, Moscow, 119991, Russia}


\date{\today}

\begin{abstract}
We study a quantum system that consists of two fermionic chains  coupled by a driven quantum point contact (QPC). The QPC contains a bond with a periodically varying tunneling amplitude. Initially the left chain is packed with fermions while the right one is empty.  We numerically track the evolution of the system and demonstrate that, at frequencies above a critical one, the current through the QPC halts, and the particle imbalance  between the chains remains forever. This implies a spectacular breakdown of the Floquet version of the  eigenstate thermalization hypothesis which predicts a homogeneous particle density profile at large times. We confirm the effect for various  driving protocols  and interparticle interactions.
\end{abstract}

\maketitle


\paragraph{Introduction.}  Leveraging a rich phenomenology of quantum transport is essential for the advancement of science and technology at nano- and microscales. The  unceasing theoretical efforts in understanding this phenomenology are of crucial importance, as they provide the foundation to interpret experimental results, predict new phenomena, and guide the design of next-generation quantum and microelectronic devices \cite{Nazarov_2009_Quantum,Platero_2004_Photon,Kohler_2005_Driven,Ryndyk_2016_Theory,Landi_2022_Nonequilibrium}.



\begin{figure}[t]
    \includegraphics[width=0.99\linewidth]{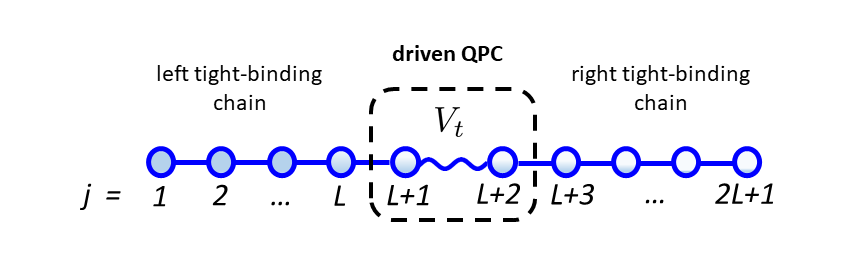}

\caption{
The system under study consists of two fermionic tight-binding chains coupled by a bond with a periodically driven tunneling amplitude. The bond  constitutes the time-dependent quantum point contact.
\label{fig system}
}
\end{figure}

Here we contribute to these efforts by studying a system of  two tight-binding fermionic chains connected by a quantum point contact (QPC) with a periodically driven tunneling amplitude, see Fig. \ref{fig system}. Initially, the left chain is filled by fermions while the right one is empty.

This system has been studied previously in  the case of noninteracting fermions \cite{gamayun2021nonequilibrium}. It has been shown  to exhibit a nonequilibrium phase transition under the variation of the driving frequency: while for low frequencies the QPC is conducting, for frequencies above the single-particle band gap the QPC turns to be insulating, and the initial imbalance of particle densities lasts forever \cite{gamayun2021nonequilibrium}.\footnote{See also ref. \cite{Znidaric_2011_Transport} where a somewhat similar nonequilibrium phase transition was found for a boundary-driven integrable spin chain. See also a recent ref. \cite{Romero-Bastida_2024_Effect} for a similar transition in a system of coupled classical harmonic oscillators.}

At first sight, the latter behavior should become impossible when the interactions between the fermions come  into play. This expectation is based on the Floquet version of the  eigenstate thermalization hypothesis (the Floquet ETH) \cite{Lazarides_2014_Equilibrium,D'Alessio_2014_Long-time}  which is believed to hold for generic interacting periodically-driven quantum many-body systems \cite{Lazarides_2014_Equilibrium,D'Alessio_2014_Long-time,Seetharam_2018_Absence,Ye_2020_Emergent,Pizzi_2020_Time,Ikeda_2021_Fermi,Morningstar_2023_Universality}.
The Floquet ETH predicts that a state locally indistinguishable from the infinite-temperature thermal state will establish in the long run, which implies the homogeneous distribution of particles across both chains.

Here we demonstrate that this expectation is wrong: the QPC remains insulating above a critical frequency even  when the interactions between the fermions are on.  This implies the breakdown of the Floquet ETH above the critical frequency.

The rest of the Letter is organized as follows. In the next section we introduce our model, observables and methods. Then we report and interpret the results. Finally, we discuss the results and give an outlook of possible further developments.

\medskip
\paragraph{Model.}
The system under study consists of two one-dimensional chains harboring spinless  fermions, see Fig.~\ref{fig system}. The lengths of the chains are $(L+1)$ and $L$. They are chosen to be unequal to avoid degeneracies  due to the reflection symmetry. The chains are connected by a driven QPC. The total Hamiltonian reads
\begin{equation}\label{H}
    H_t = H_L+H_R+V_{\text{int}} +V_t,
\end{equation}
Here $H_L$ and $H_R$ describe hopping along left and right chains, respectively,
\begin{align}\label{HL}
H_L=&-\frac{1}{2}\sum_{j=1}^{L}( c^{\dagger}_jc_{j+1}+ c^{\dagger}_{j+1}c_j),\\
H_R=&-\frac{1}{2}\sum_{j=L+2}^{2L}( c^{\dagger}_jc_{j+1}+ c^{\dagger}_{j+1}c_j),
\label{HR}
\end{align}
$c_j$ are fermionic annihilation operators, $V_{\text{int}}$ is the interparticle interaction term and $V_t$ is the time-periodic Hamiltonian of the QPC acting exclusively on the pair of boundary sites, see Fig \ref{fig system}. Here we focus on the specific  $V_t$ and $V_{\text{int}}$,
\begin{align}
    V_t& =  -\frac{1}{2} \,f_t\, \left( \hat{c}^\dagger_{L+1}\hat{c}_{L+2}+ \hat{c}^\dagger_{L+2}\hat{c}_{L+1}\right),\qquad f_t=\sin\omega t,\label{Vt}\\
V_{\text{int}} & =  U\sum_{j=1}^{2L}n_jn_{j+1},\label{Vint}
\end{align}
where $\omega$ is the driving frequency, $n_j = c^{\dagger}_j c_j$ is the particle number operator  at the $j$'th site and  $U$  is the interaction strength. Other types of $V_t$ and $V_{\text{int}}$ are addressed in the Supplementary Material \cite{supp}. The noninteracting case of $U=0$ was studied in ref. \cite{gamayun2021nonequilibrium}.

The number of fermions is conserved and we fix it to be $N=L+1$. Initially, the fermions fill the left chain, while the right chain is empty: \begin{equation}
\Psi_0=\left(\prod_{j=1}^{L+1} c_j^\dagger\right) |\textrm{vac}\rangle.
\end{equation}
The state of the system $\Psi_t$ evolves according to the Schr\"odinger equation  $i \partial_t \Psi_t=H_t \Psi_t$.

We focus on a specific observable, the number of fermions in the right chain:
\begin{equation}
N_R=\sum_{j=L+2}^{2L+1}n_j.
\end{equation}
Its expectation value $\langle N_R\rangle_t=\langle\Psi_t|N_R|\Psi_t\rangle$ is initially zero and grows as a result of particle flow through the QPC.

\begin{figure}[t]
    \includegraphics[width=1.1\linewidth]{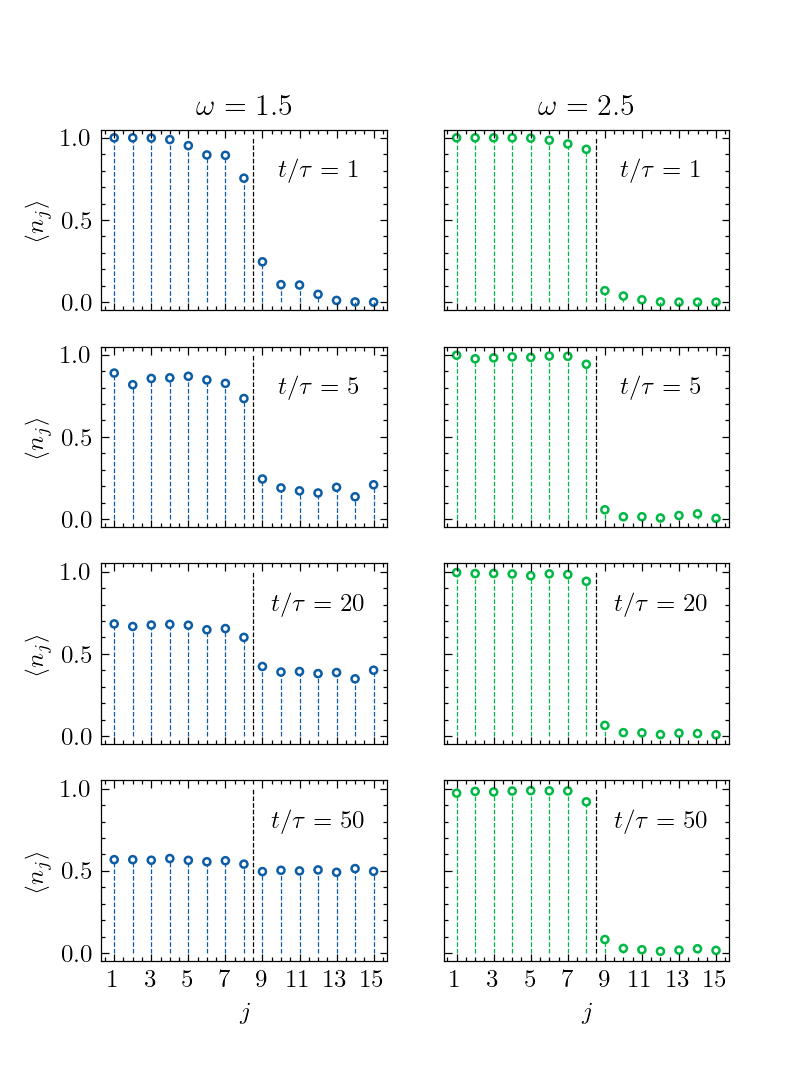}
\caption{Evolution of particle density for two driving frequencies. For a frequency below the critical one (left column) particles tend to distribute uniformly over sites. For a frequency above the critical one (right column)  the particle imbalance between the chains is retained. Dashed horizontal line delineates two chains. \label{fig: distribution}}
\end{figure}

At large times $N_R$ is expected to fluctuate around its long-time average $\langle N_R\rangle_\infty \equiv \lim\limits_{t\rightarrow \infty} t^{-1} \int_0^t dt' \langle N_R\rangle_{t'}$. In the absence of degeneracies in the Floquet spectrum,  $\langle N_R\rangle_\infty$ can be computed as
\begin{equation}
\label{eq:steady}
    \langle N_R\rangle_\infty=
\sum_\alpha
|\langle\Phi_{\alpha}|\Psi_0\rangle|^2
\langle \Phi_{\alpha}|N_R|\Phi_{\alpha}\rangle,
\end{equation}
where $\Phi_{\alpha}$ are the Floquet eigenstates, i.e. eigenstates of the operator of quantum  evolution over the driving period $\tau=2\pi/\omega$.

\begin{figure*}[ht]
    \includegraphics[width=0.48\textwidth]{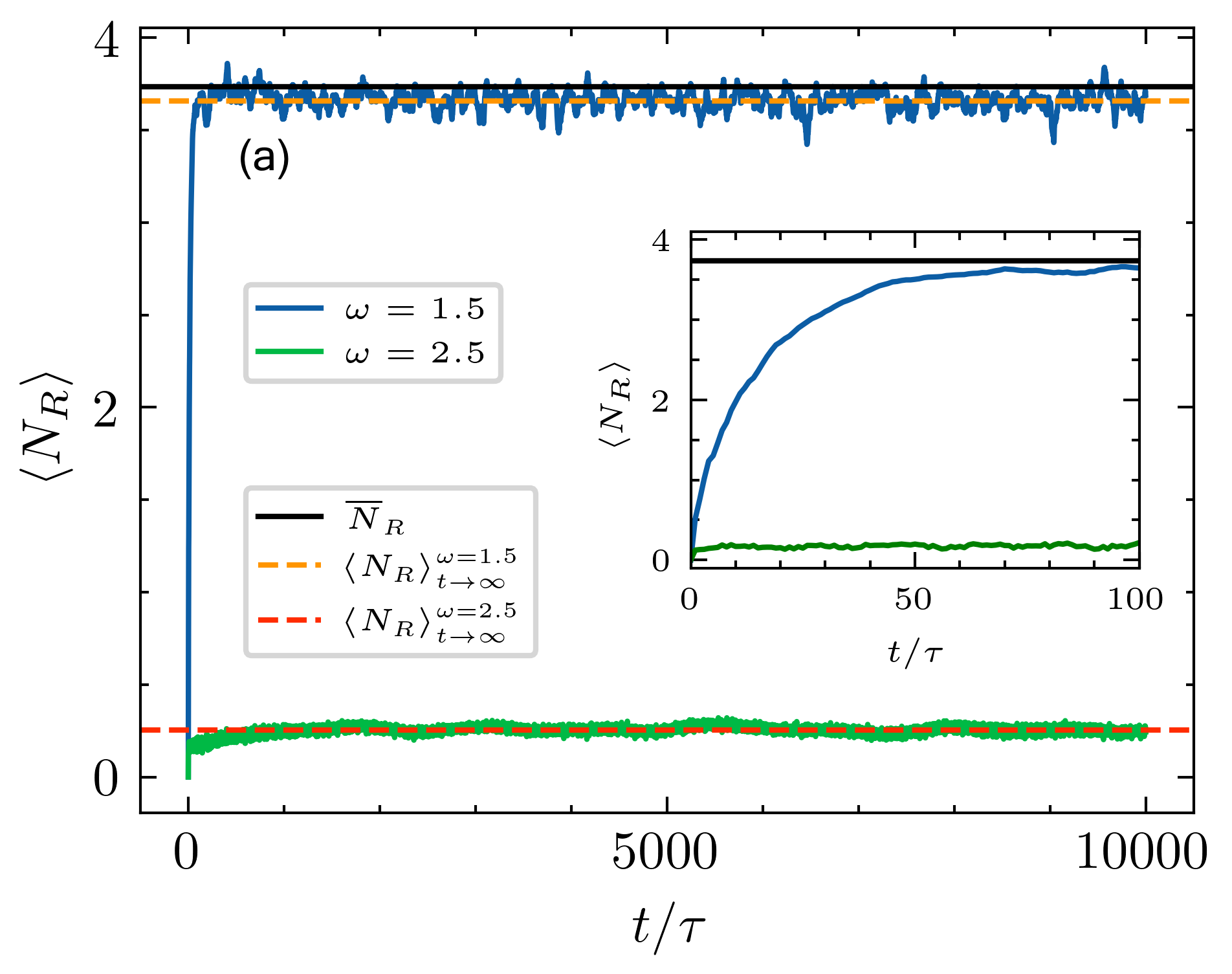}
    \includegraphics[width=0.49\textwidth]{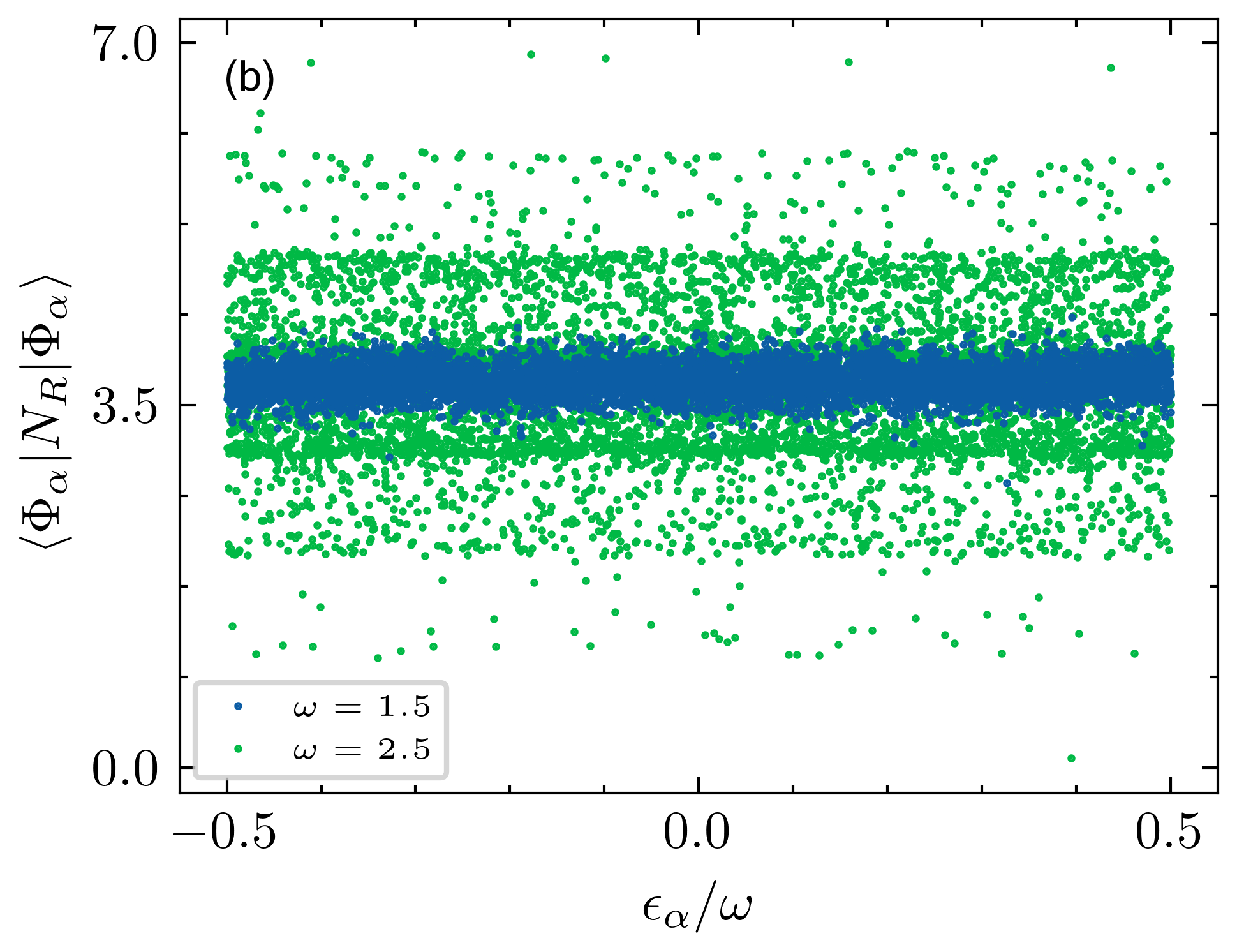}
\caption{(a) The number of particles in the right chain, $\langle N_R\rangle_t$, as a function of time for frequencies below ($\omega = 1.5$) and above ($\omega = 2.5$) the critical frequency. Steady state values computed according to eq.~(\ref{eq:steady}) are shown by horizontal dashed lines. The horizontal black line shows the value of $N_R$ corresponding to the uniform distribution of particles over the chains.  The interaction strength is $U=0.5$. The total number of sites and fermions is  $2L+1 =15$ and $N=8$, respectively.
(b) Diagonal matrix elements of the operator $N_R$ in Floquet basis for different Floquet energies $\epsilon_{\alpha}$.
\label{fig:1}}
\end{figure*}

\begin{figure*}[ht]
    \includegraphics[width=0.48\textwidth]{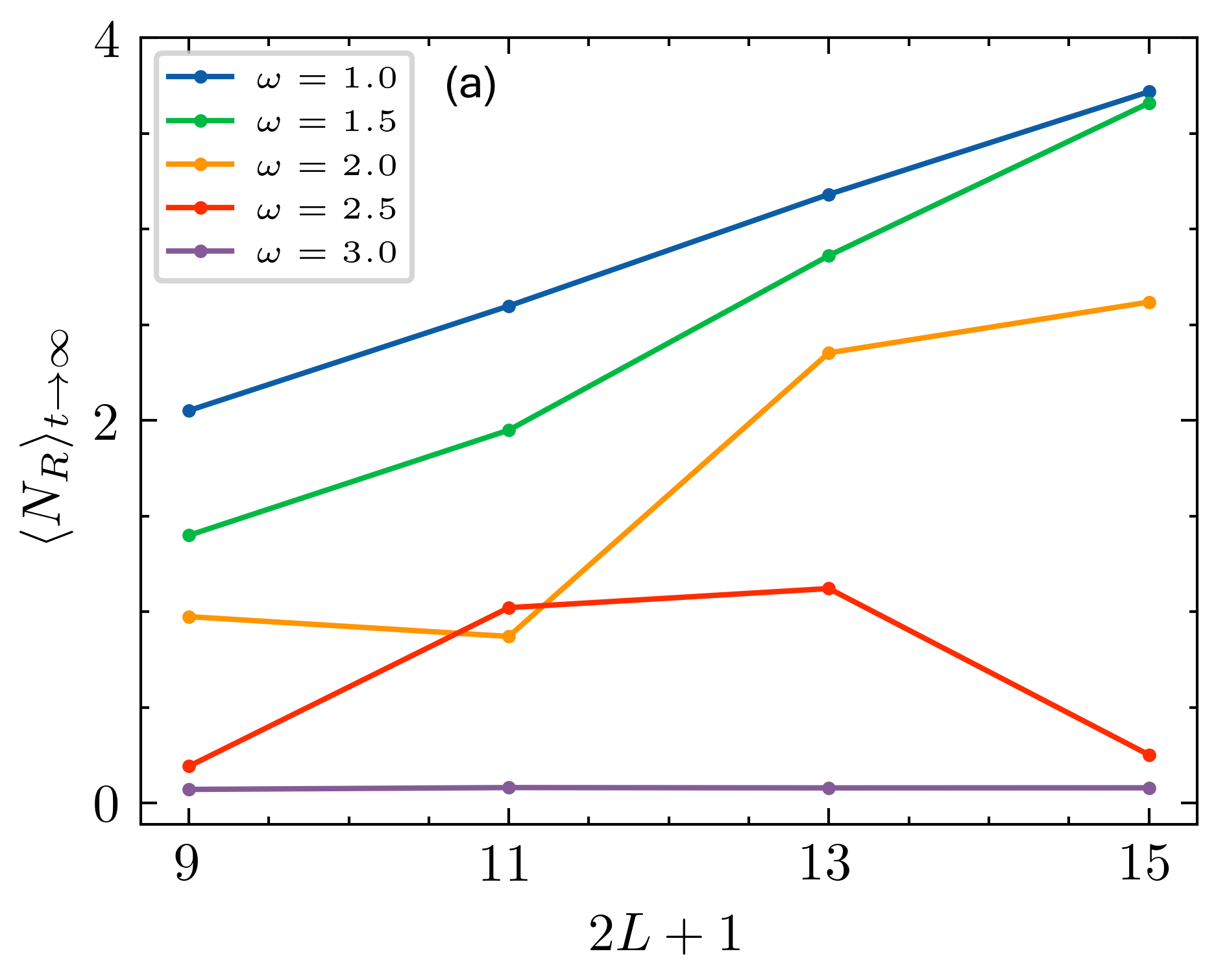}
\includegraphics[width=0.5\textwidth]{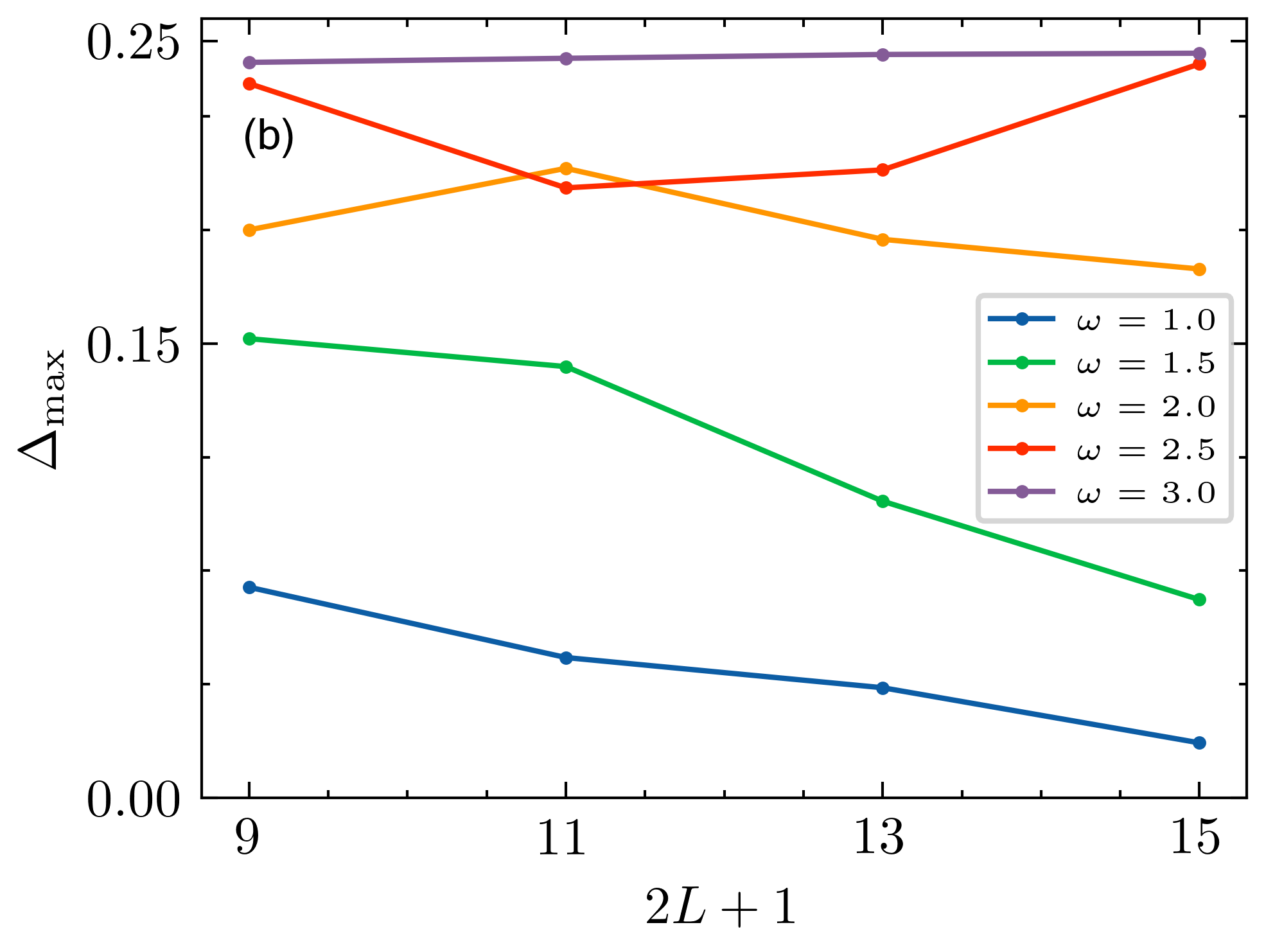}
 \caption{The scaling with the system size of (a) the steady state value $\langle N_R\rangle_\infty$ and (b) the maximal deviation of the diagonal matrix element $\langle \Phi_\alpha |N_R| \Phi_\alpha \rangle$ from the uniform value $\overline N_R $. The interaction strength is $U=0.5$. One can see the quantitative difference in the scaling behavior for frequencies below and above the critical frequency $\omega_c\simeq2$.}
\label{fig:2}
\end{figure*}

\medskip
\paragraph{Results.} We use the QuSpin package \cite{Weinberg_2017_QuSpin,Weinberg_2019_QuSpin} to  numerically solve the Schr\"odinger equation and compute Floquet spectrum and eigenstates for finite systems of sizes up to $2L+1=15$. We do this for various frequencies. We find  qualitative differences in the dynamics and Floquet eigenstates  above and below some critical frequency $\omega_c$, see Fig. \ref{fig:1}. For not too large  interaction strengths $U\lesssim 1$, this critical frequency approximately equals $2$, as in the noninteracting case \cite{gamayun2021nonequilibrium}.

For frequencies below the critical one  the system evolves towards thermalization: particles flow from the left chain to the right one through the QPC and eventually distribute uniformly across both chains, see Fig. \ref{fig: distribution} and Fig. \ref{fig:1} (a). The equilibrium value $\langle N_R\rangle_\infty$ coincides, up to small finite size corrections, with the value
\begin{equation}
\overline N_R = N \,L/(2L+1)
\end{equation}
obtained for the uniform distribution of particles over the sites of the system.

In contrast, for frequencies above the critical one the QPC effectively turns insulating: the particles stay in the left chain and never come to the right one. In this case the equilibrium value $\langle N_R\rangle_\infty$ is around zero, as shown in  Fig. \ref{fig: distribution} and Fig. \ref{fig:1} (a).

The  diagonal matrix elements $\langle \Phi_{\alpha}|N_R|\Phi_{\alpha}\rangle$ of $N_R$ over the Floquet basis is shown in Fig. \ref{fig:1} (b). The patterns below and above the critical frequency are distinctly different. Below the critical frequency the matrix elements tightly concentrate around $\overline N_R$, consistent with the Floquet ETH.

In contrast, above the critical frequency matrix elements are scattered over a wide interval, with minimum and maximal instances almost reaching minimal and maximal eigenvalues of $N_R$, which are $0$ and $L$, respectively.

To check whether our results can be  relevant in the thermodynamic limit,  we perform a finite size scaling, see Fig. \ref{fig:2}. In Fig. \ref{fig:2} (a) the scaling of the equilibrium value $\langle N_R\rangle_\infty$ is shown. One can see that for frequencies below the critical one it scales linearly with the system size, as expected under the assumption of uniform  particle distribution in equilibrium. In contrast, for frequencies above the critical one no growth is observed.

In Fig. \ref{fig:2} (b) the Floquet ETH is tested directly. There we plot the maximal deviation $\Delta_{\max}$ of the diagonal matrix element $\langle \Phi_\alpha |N_R| \Phi_\alpha \rangle$ from the uniform value $\overline N_R $ (maximization is performed over all Floquet eigenstates), normalized to the system size:
\begin{equation}
\Delta_{\max}=(2L+1)^{-1} \,\max_\alpha \Big| \langle \Phi_\alpha |N_R| \Phi_\alpha \rangle - \overline N_R \Big|.
\end{equation}
If the Floquet ETH is valid, $\Delta_{\max}$ should decrease with the system size. This is indeed the case for frequencies below the critical one, as can be seen from  \ref{fig:2} (b). In contrast, above the critical frequency this figure of merit stays approximately constant, indicting the breakdown of the Floquet ETH.

We have verified that the above qualitative picture is robust with respect to variations of the interaction term $V_{\text{int}}$, as well as the periodic function $f_t$ that modulates the hopping amplitude in the QPC in eq. \eqref{Vt}, provided it averages to zero over the period, $\int_0^{\tau}f_t\,dt=0$. The results for a number of exemplary models are presented in the Supplement \cite{supp}.

\medskip
\paragraph{Discussion and outlook.} To summarize, we have demonstrated a nonequilibrium phase transition between thermal and athermal phases in a periodically driven interacting disorder-free quantum many-body system. The system consists of two fermionic chains connected by a driven QPC. The transition occurs upon the variation of the driving frequency. For frequencies below the critical one the  QPC is conducting, and the fermions eventually spread homogeneously across the chains, in accordance with the Floquet ETH. For frequencies exceeding the critical frequency, the QPC is insulating, and the initial particle imbalance between the chains remains forever, in stark contrast to the prediction based on the Floquet ETH.

The above frequency dependence of the effect sharply distinguishes it from other known types of the Floquet thermalization breakdown. In particular, in non-generic Floquet systems featuring integrability, exact quantum many-body scars or Hilbert space fragmentation, the violation of the Floquet ETH and the thermalization breakdown typically take place for arbitrary driving frequencies, and athermal eigenstates often are independent on the frequency \cite{Gritsev_2017_Integrable,Mizuta_2020_Exact,Yarloo_2020_Homogeneous,Khemani_2020_Localization,Mukherjee_2020_Collapse,Moudgalya_2022_Quantum,teretenkov2023exact}. Alternatively, the breakdown or slowdown of thermalization can occur for a discrete set of frequencies \cite{Dunlap_1986_Dynamic,Dunlap_1988_Dynamic,Grossmann_1991_Coherent,Eckardt_2009_Exploring,Das_2010_Exotic,Bhattacharyya_2012_Transverse,Luitz_2017_Absence,Agarwala_2017_Effect,
Haldar_2018_Onset,Haldar_2021_Dynamical,Fleckenstein_2021_Prethermalization,Sugiura_2021_Many-body,Aditya_2023_Dynamical,Ghosh_2023_Prethermal,
guo2024dynamical}. A completely different frequency dependence of the Floquet ETH  violation reported here may indicate a distinct origin of this violation -- a question left for further studies.

It should be emphasized that the thermalization breakdown reported here can not be explained by the Floquet prethermalization \cite{Abanin_2015_Exponentially,Mori_2016_Rigorous,Rubio-Abadal_2020_Floquet,Beatrez_2021_Floquet,Ho_2023_Quantum}. The latter phenomenon implies the thermalization slowdown at large frequencies, with the thermalization time depending exponentially on the frequency.
If the system under study exhibited ``mere'' prethermalization, the particle imbalance between the chains would be eventually washed out. We know that this is not the case, since in finite systems under study we have access to arbitrary long times, as well as to the infinite-time averages \eqref{eq:steady}. Furthermore,  the critical frequency is not, in fact, large enough for the prethermalization mechanism to be effective, as we demonstrate in the Supplement \cite{supp}.


An intriguing question for further studies concerns the scope of the reported effect. While we have demonstrated its robustness in one-dimensional geometry, our numerical tests do not extend to higher dimensions. It is not uncommon when exotic out-of-equilibrium phenomena in one dimension have no  counterparts in higher dimensions \cite{Zvonarev_2007_Spin,Knap_2014_Quantum,Burovski_2014_Momentum,Lychkovskiy_2014_Perpetual,Lychkovskiy_2015_Perpetual,Bertini_2021_Finite-temperature}. It is therefore reasonable to ask whether the reported type of the Floquet thermalization breakdown  extends beyond one-dimensional geometry. We note that this question, while challenging for numerical exploration, can perhaps be addressed with the state-of-the-art quantum simulators with planar  arrangement of qubits \cite{Arute_2019_Quantum,Gong_2021_Quantum,Ebadi_2021_Quantum,Scholl_2021_Quantum,Lienhard_2018_Observing}.

As a word of caution, we note that, in the case of noninteracting fermions, dissipation (caused e.g. by phonon-fermion interactions) destroys the phase transition and makes the QPC conducting at any frequency \cite{Ermakov_2024_Effect}. Likely it will have the same effect in the interacting case. This should be accounted for while conceiving experimental tests of the effect.

\begin{acknowledgments}
\paragraph{Acknowledgments.} We are grateful to  Oleksandr Gamayun and Vincenzo Alba for extremely helpful discussions and input at the initial stage of this study.   A part of this work concerning numerical calculation of particle number as a function of time (with the results presented in Figs. 2 and 3 (a))  was supported by the Ministry of Science and Higher Education of the Russian Federation (project no. 075-15-2020-788).
\end{acknowledgments}

\bibliography{bib}

\clearpage

\onecolumngrid

\renewcommand{\theequation}{S\arabic{equation}}
\setcounter{equation}{0}

\renewcommand{\thefigure}{S\arabic{figure}}
\setcounter{figure}{0}

\appendix

\section{
	{\large Supplementary material}
	\\
	 to the Letter \\
	``Fermionic transport through a driven quantum point contact:  \\
breakdown of Floquet thermalization
beyond a critical driving frequency''\\
	  by Ivan V. Dudinets  and Oleg Lychkovskiy.
}


\section{S1. Various driving protocols}

\begin{figure*}[h!]
\includegraphics[width=0.48\textwidth]{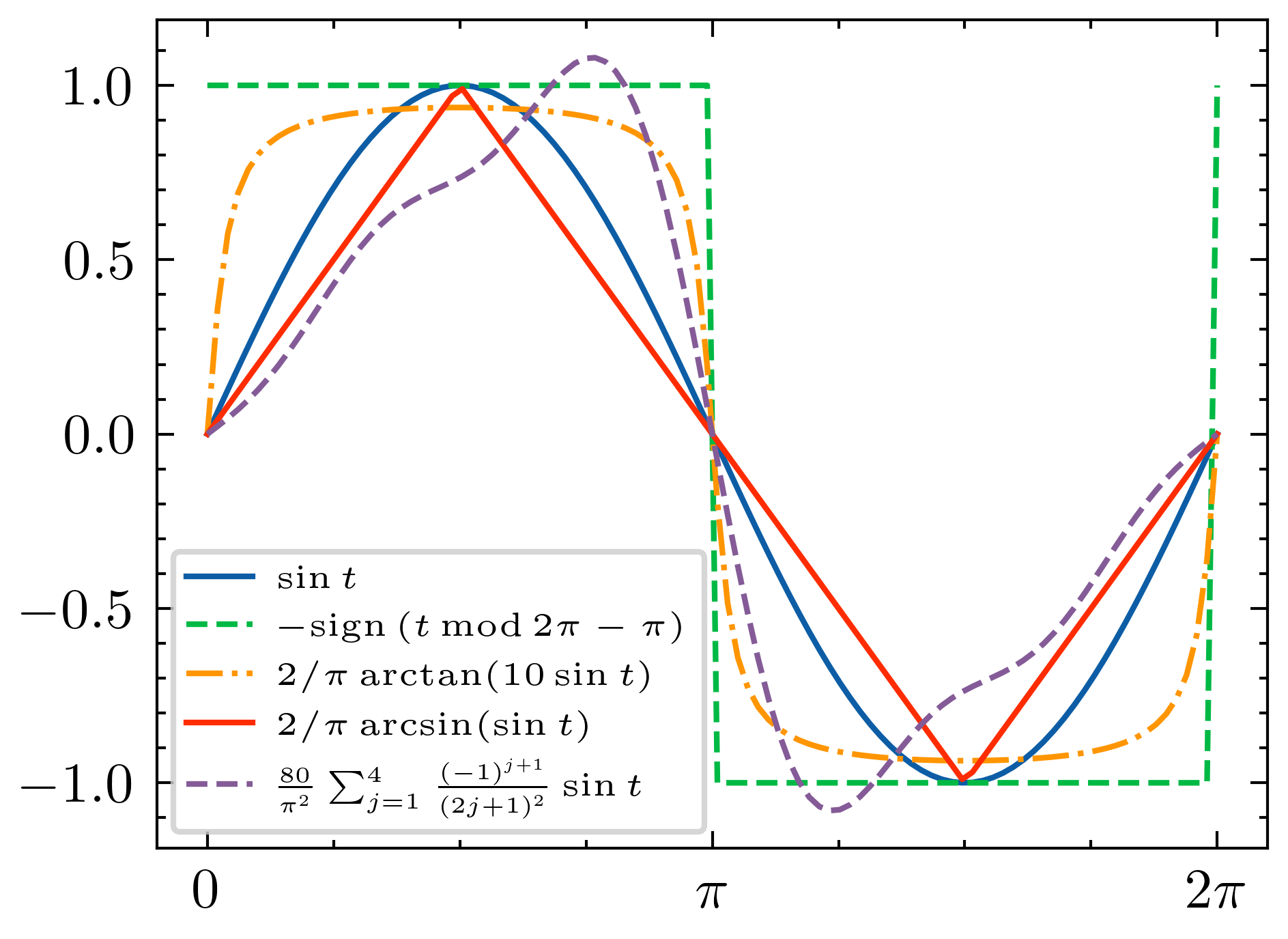}
\caption{Plot of the driving protocols, $f_t$, as a function of time.  Here  the driving frequency is set to 1.
\label{fig:ft}}
\end{figure*}

\begin{figure*}[p]
\includegraphics[width=0.48\textwidth]{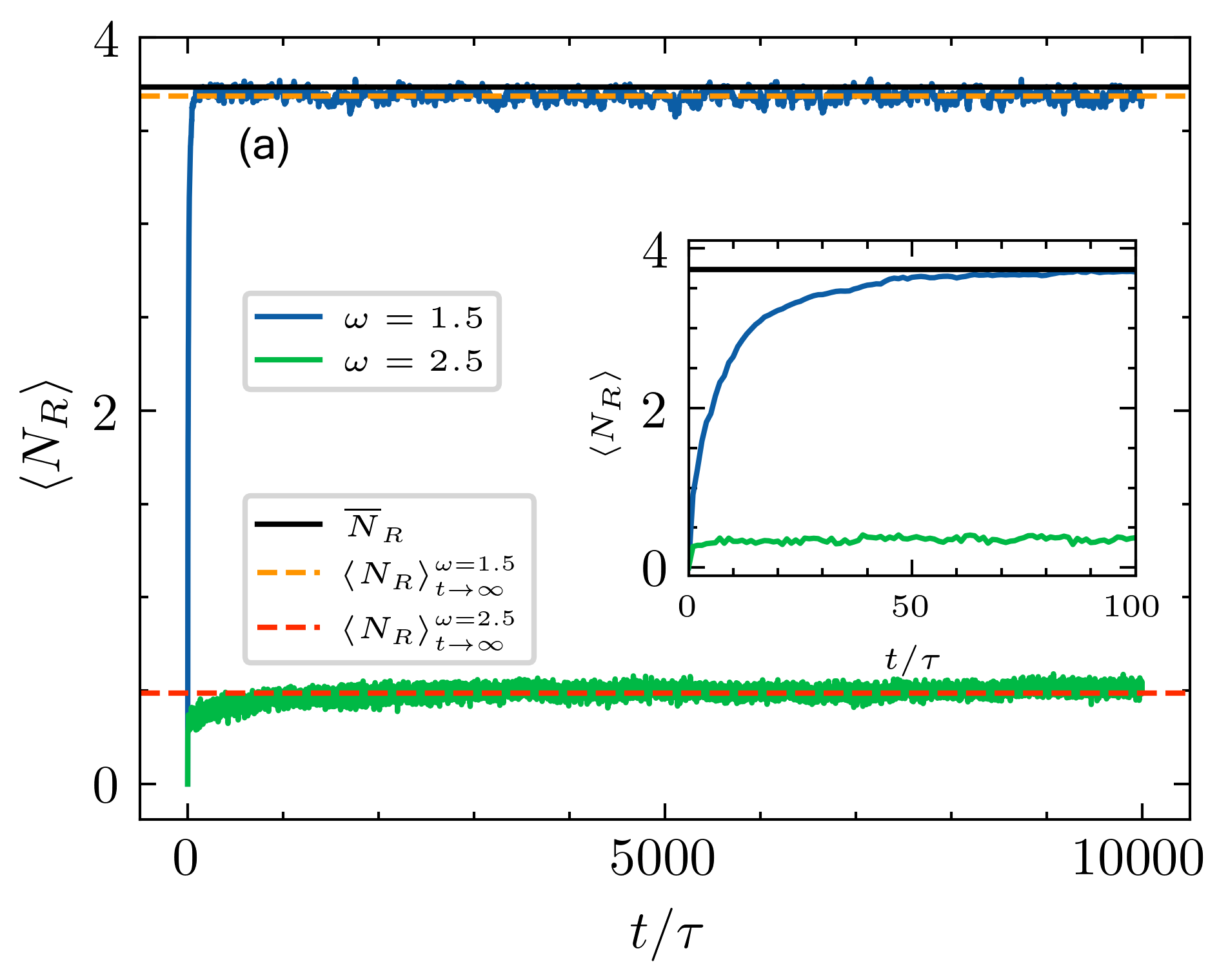}
    \includegraphics[width=0.49\textwidth]{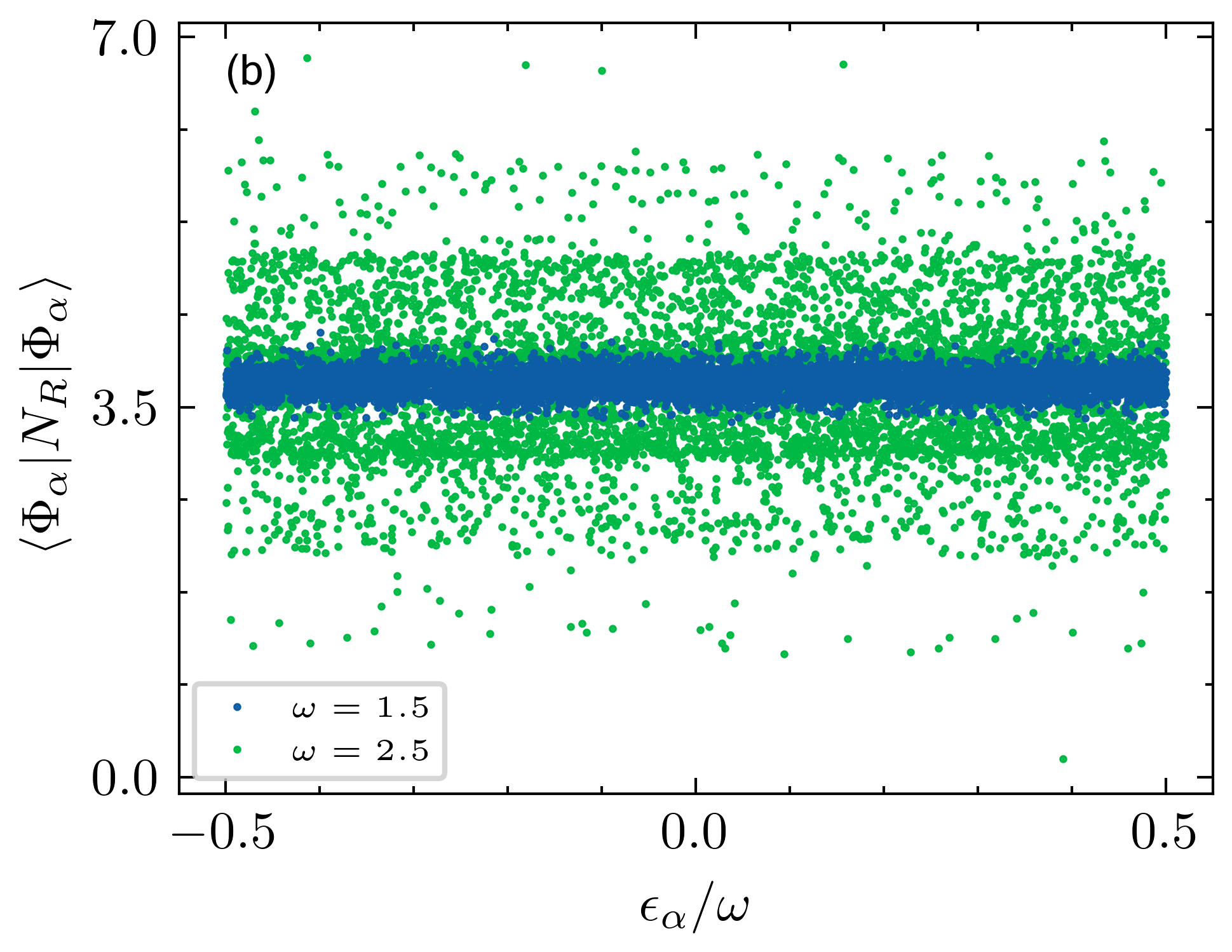}
    \\
    \includegraphics[width=0.48\textwidth]{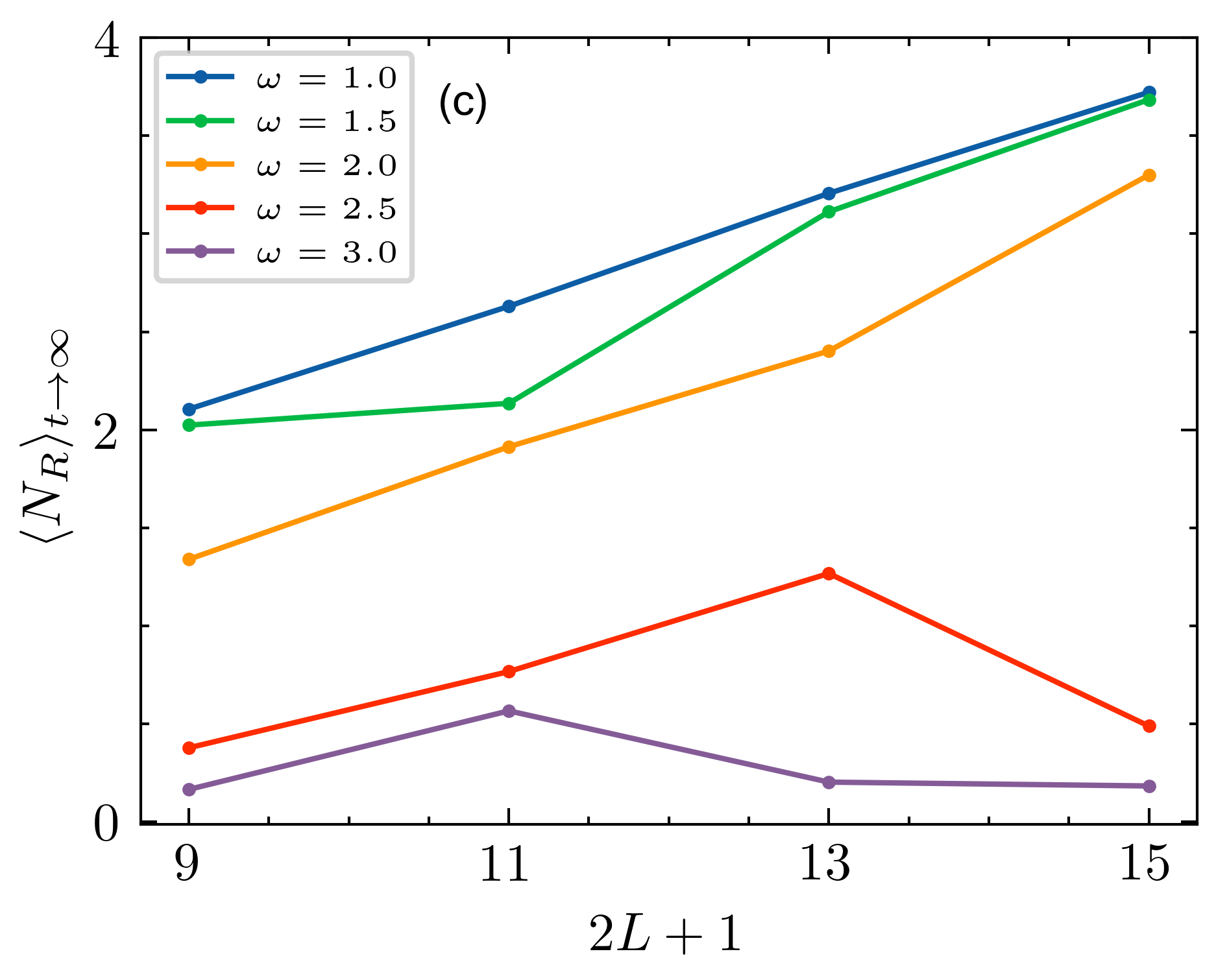}
    \includegraphics[width=0.49\textwidth]{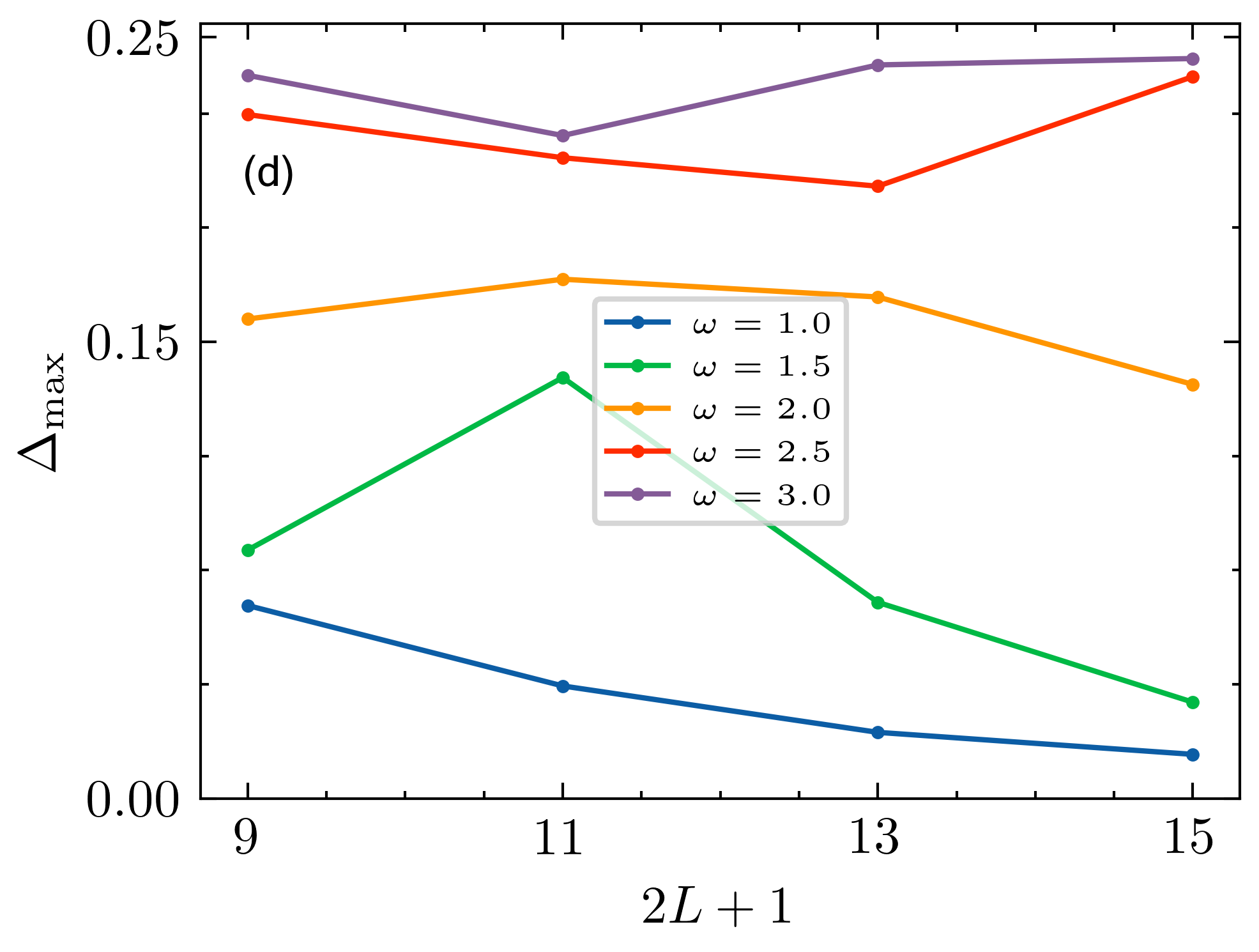}
    \caption{\label{fig:alternating} Results for the driving protocol $f_t = - \text{sign}\left(t\,\text{mod} \,\frac{2\pi}{\omega}-\frac{\pi}{\omega}\right)$. The interaction strength is $U=0.5$. $(a)$ The number of particles in the right chain, $\langle N_R\rangle_t$, as a function of time for frequencies below ($\omega = 1.5$) and above ($\omega = 2.5$) the critical frequency $\omega_c\simeq2$. Horizontal dashed lines correspond to steady state values computed according to eq.~(\ref{eq:steady}).  The horizontal black line shows the value of $N_R$ corresponding to the uniform distribution of particles over the chains.
$(b)$ Diagonal matrix elements of the operator $N_R$ in Floquet basis for different Floquet energies $\epsilon_{\alpha}$. In Figs. $(a)$ and $(b)$ the system comprises $2L+1 =15$ sites and $N=8$ fermions.
$(c)$ The steady state value $\langle N_R\rangle_\infty$ for the system of various sizes. $(d)$ The maximal deviation of the diagonal matrix element $\langle \Phi_\alpha |N_R| \Phi_\alpha \rangle$ from the uniform value $\overline N_R $.
}
\end{figure*}

\begin{figure*}[p]
\includegraphics[width=0.48\textwidth]{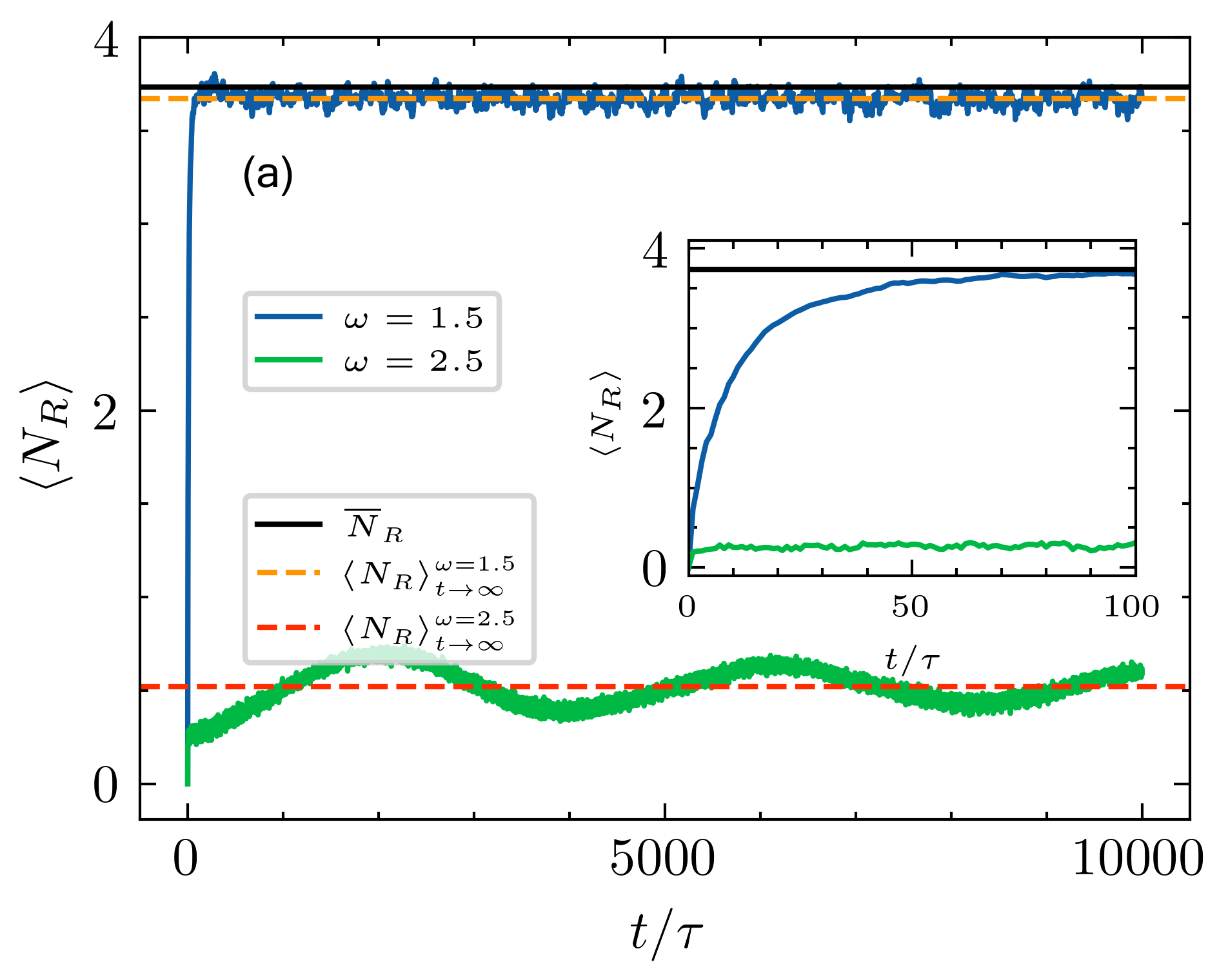}
    \includegraphics[width=0.49\textwidth]{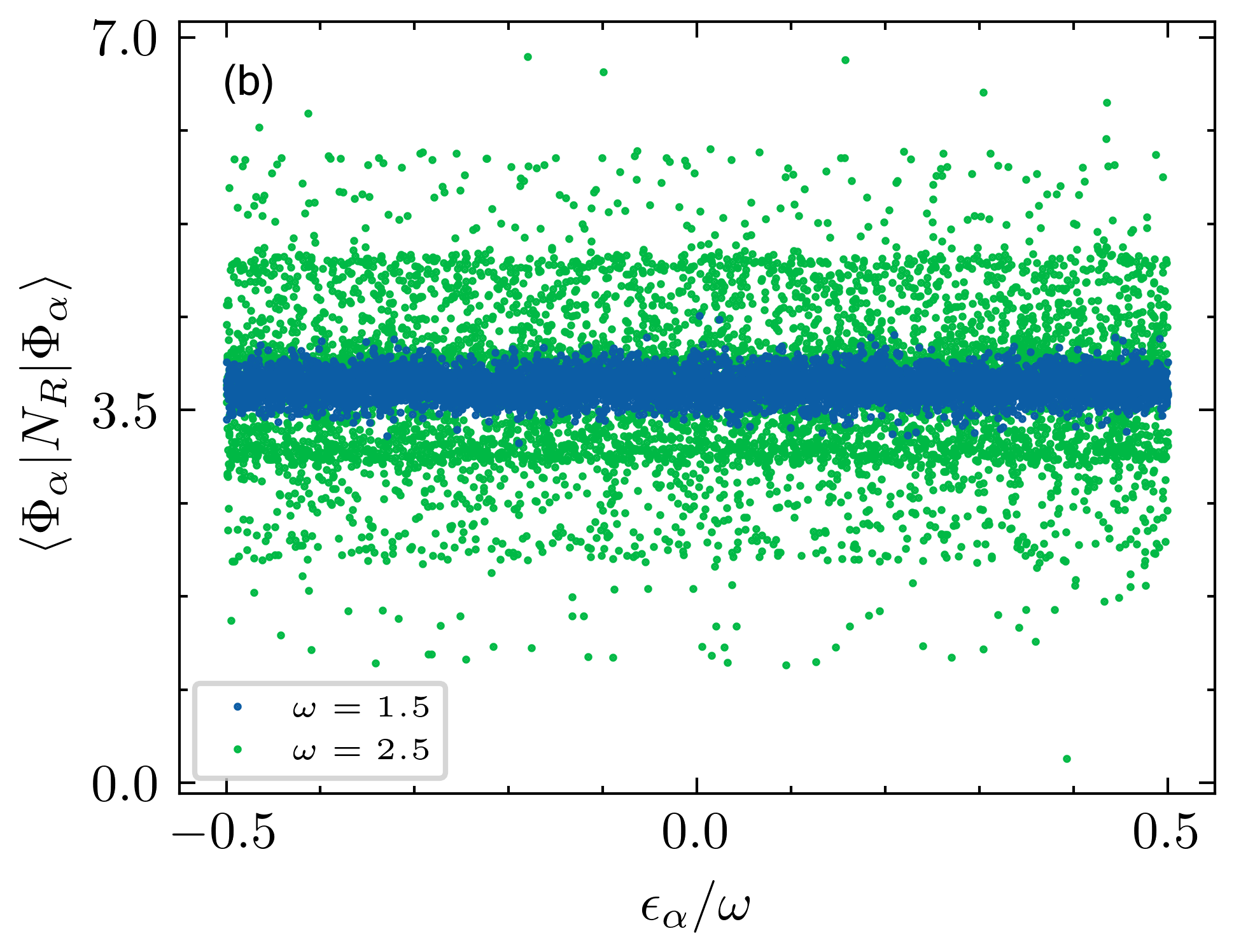}
    \\
    \includegraphics[width=0.48\textwidth]{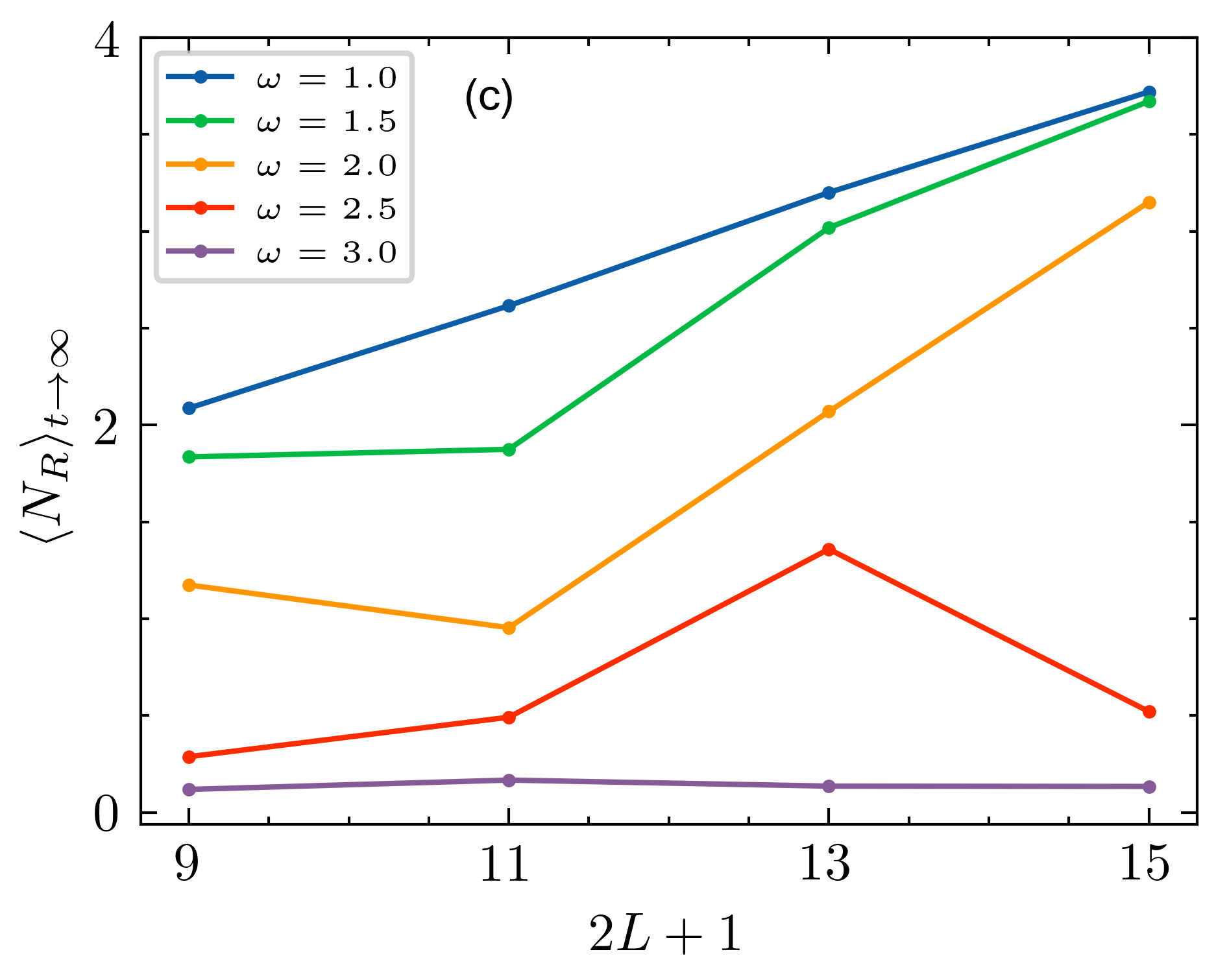}
    \includegraphics[width=0.49\textwidth]{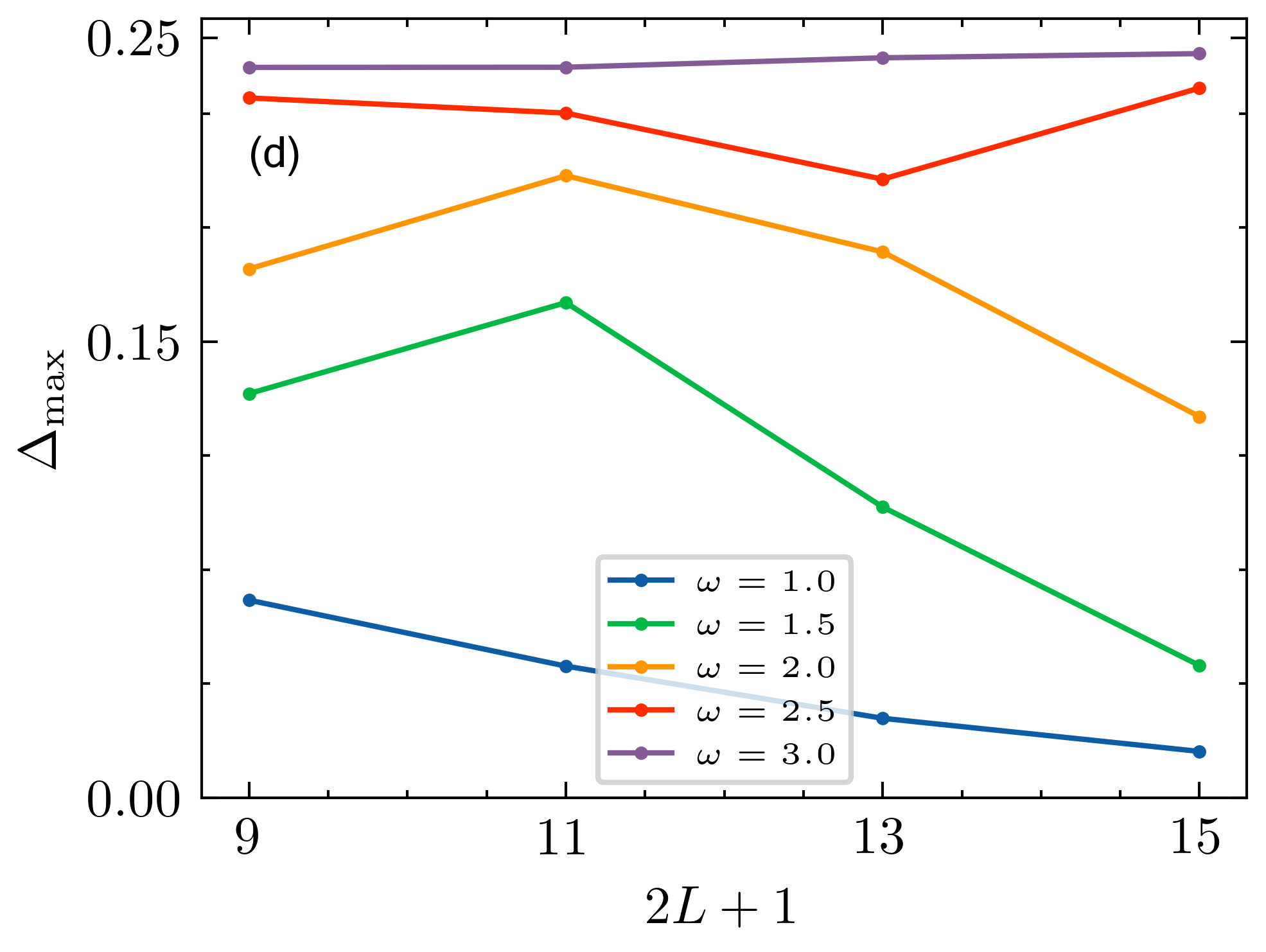}
    \caption{\label{fig:arctunneling}The same as in  Fig.~\ref{fig:alternating} except for the driving protocol $f_t = \frac{2}{\pi}\arctan\left(\sin{10 \,\omega t }\right)$.}
\end{figure*}

\begin{figure*}[ht!]

\includegraphics[width=0.48\textwidth]{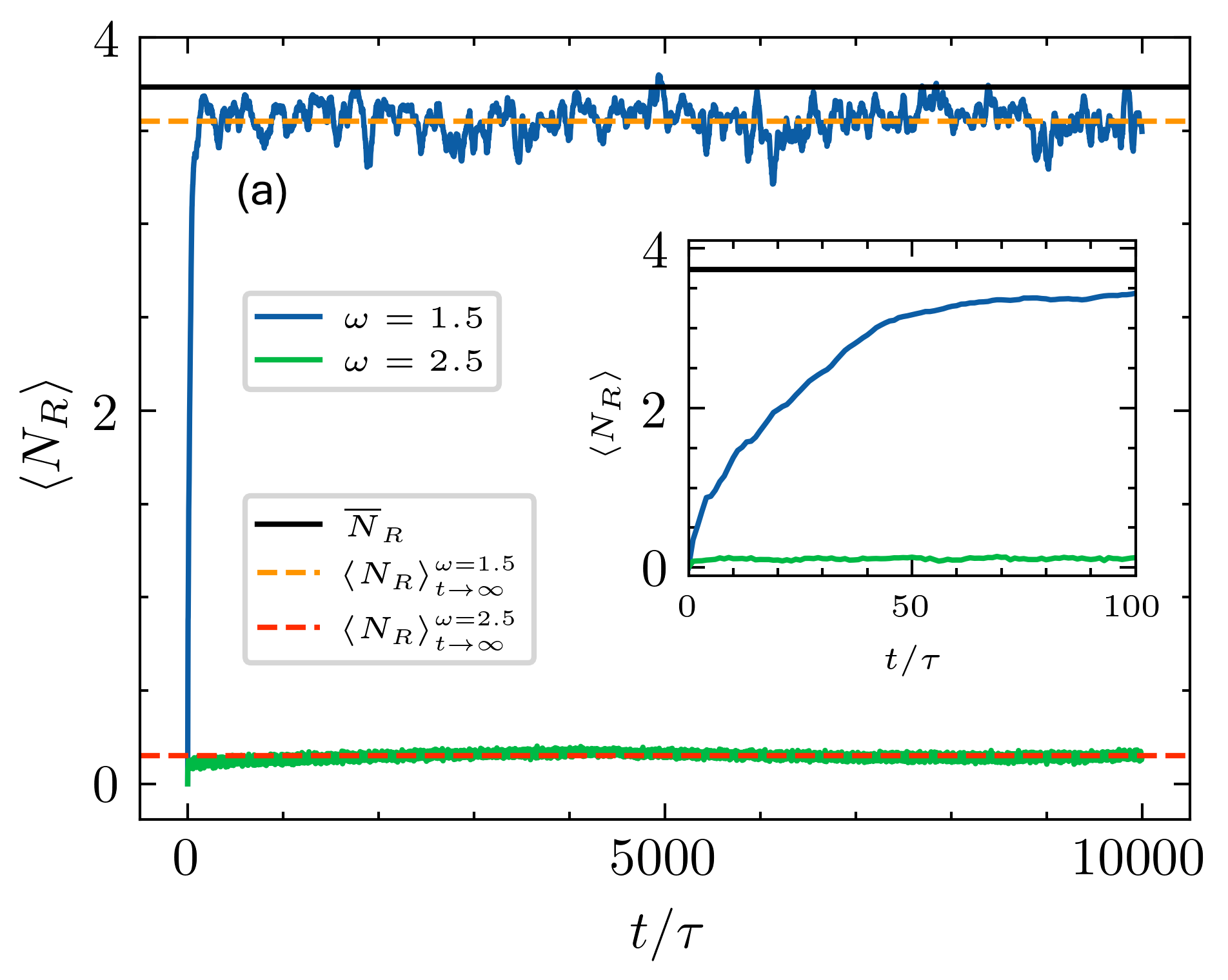}
    \includegraphics[width=0.49\textwidth]{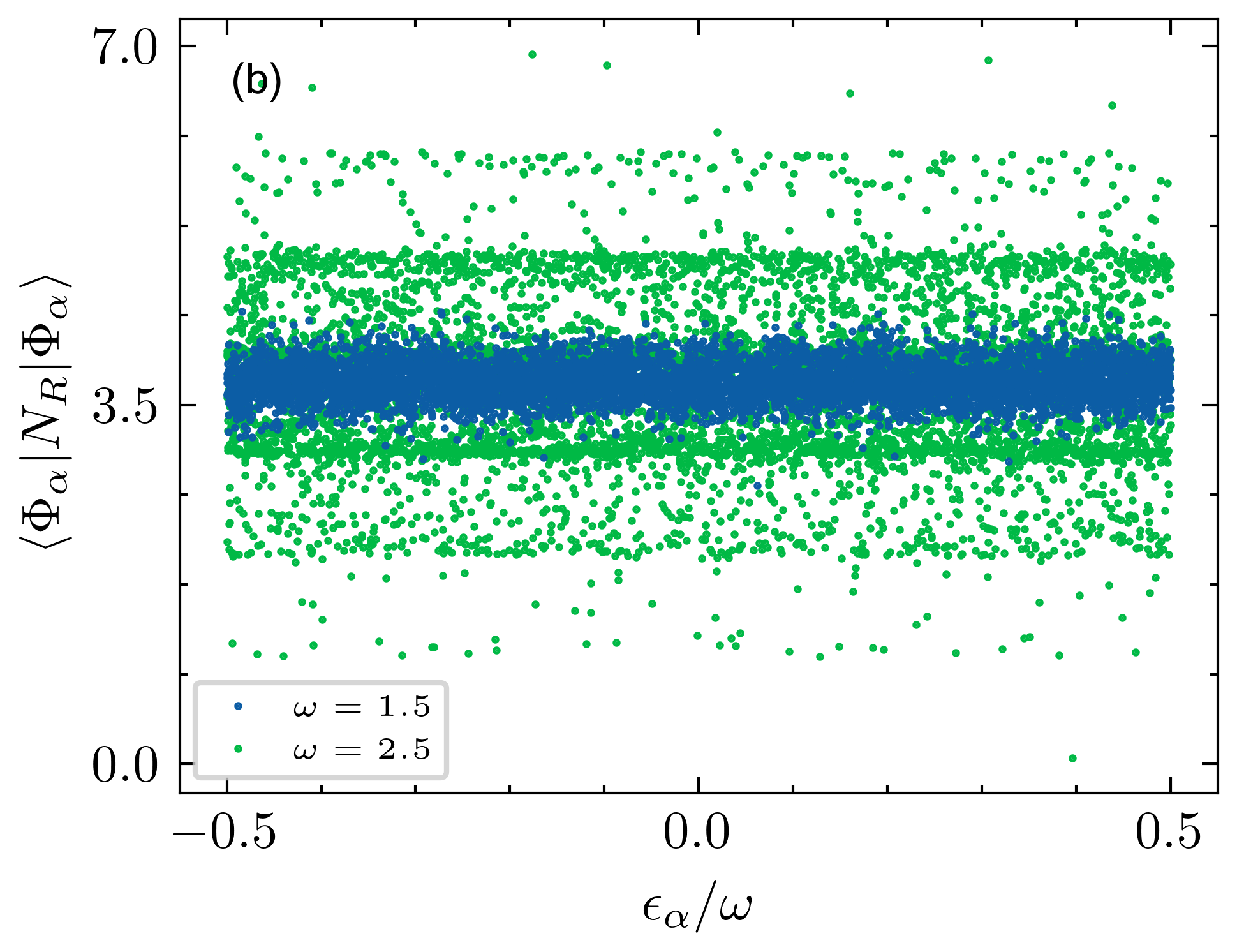}
    \\
    \includegraphics[width=0.48\textwidth]{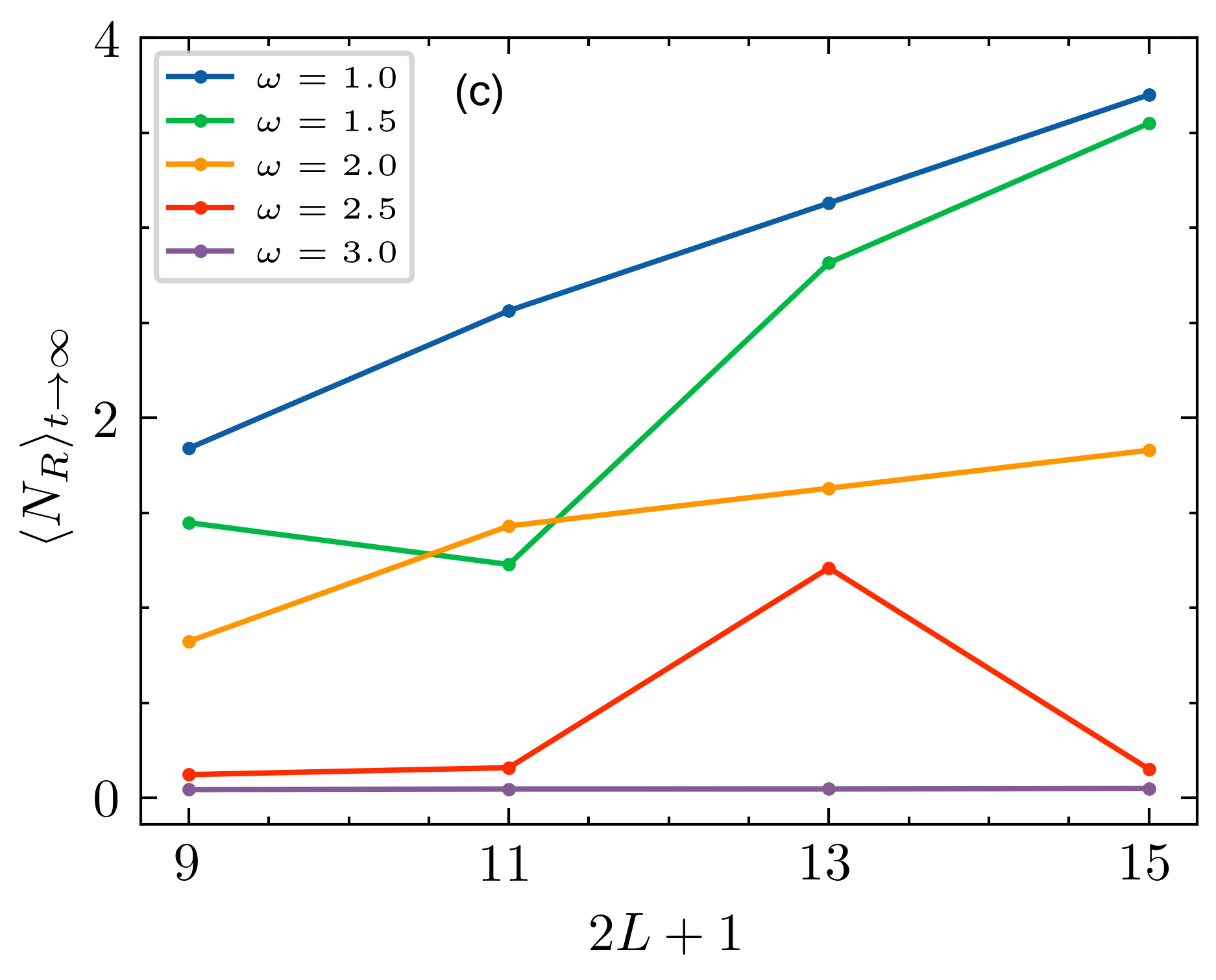}
    \includegraphics[width=0.49\textwidth]{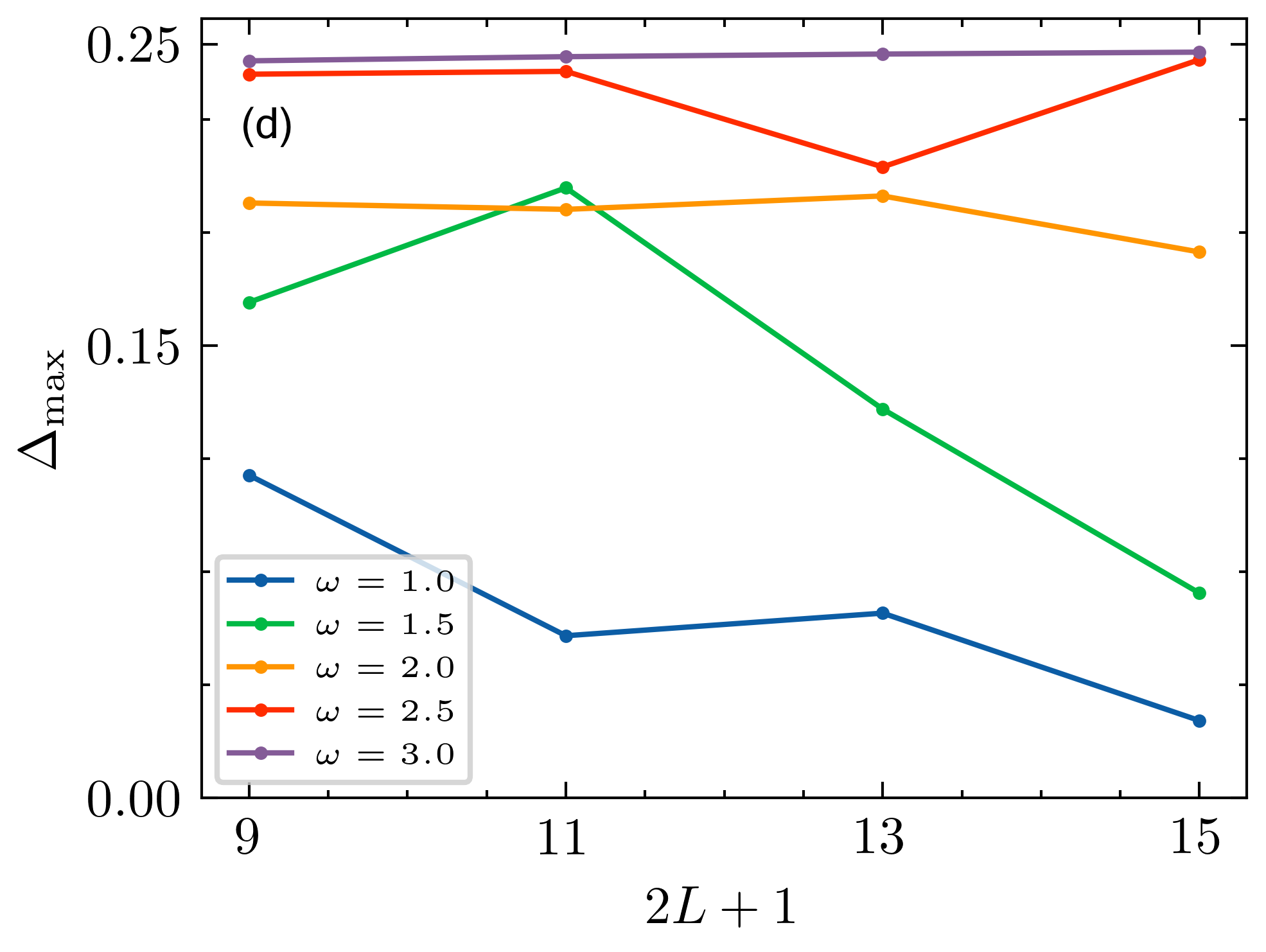}
    \caption{\label{fig:arcsinsin}The same as in  Fig.~\ref{fig:alternating} except for the driving protocol $f_t = \frac{2}{\pi}\arcsin\left(\sin{\omega t }\right)$.}
\end{figure*}

\begin{figure*}[ht]
    \includegraphics[width=0.48\textwidth]{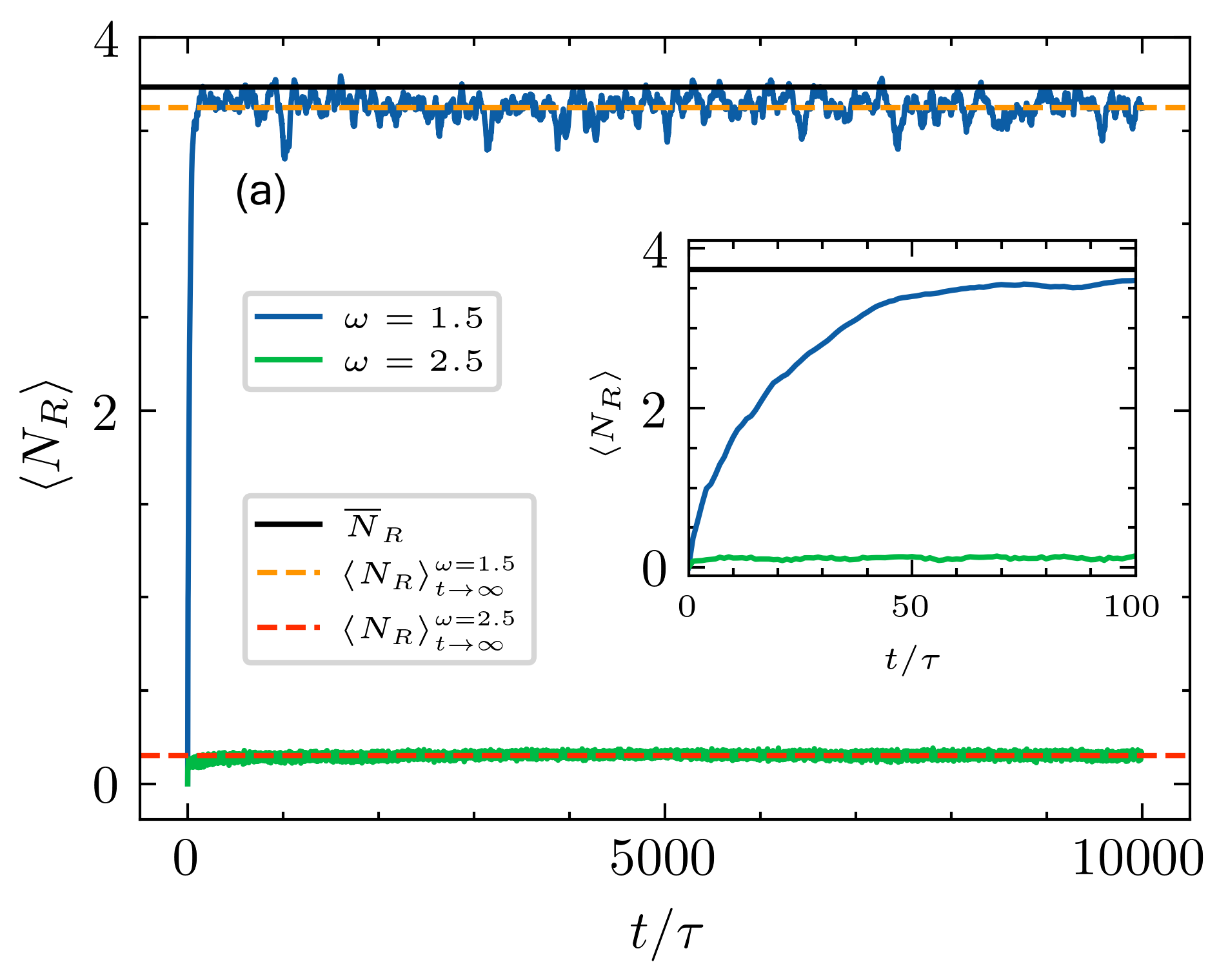}
    \includegraphics[width=0.49\textwidth]{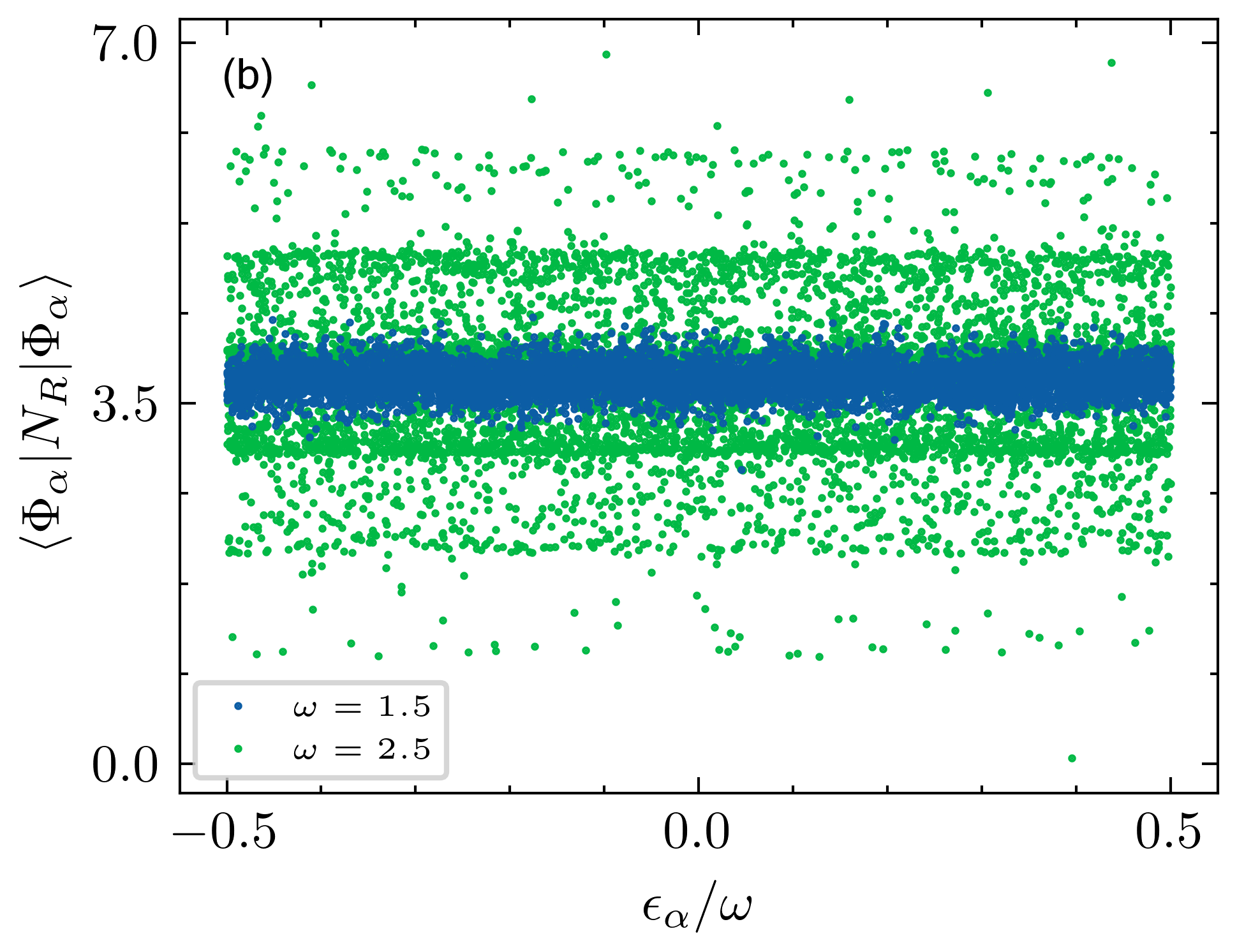}
    \\
\includegraphics[width=0.48\textwidth]{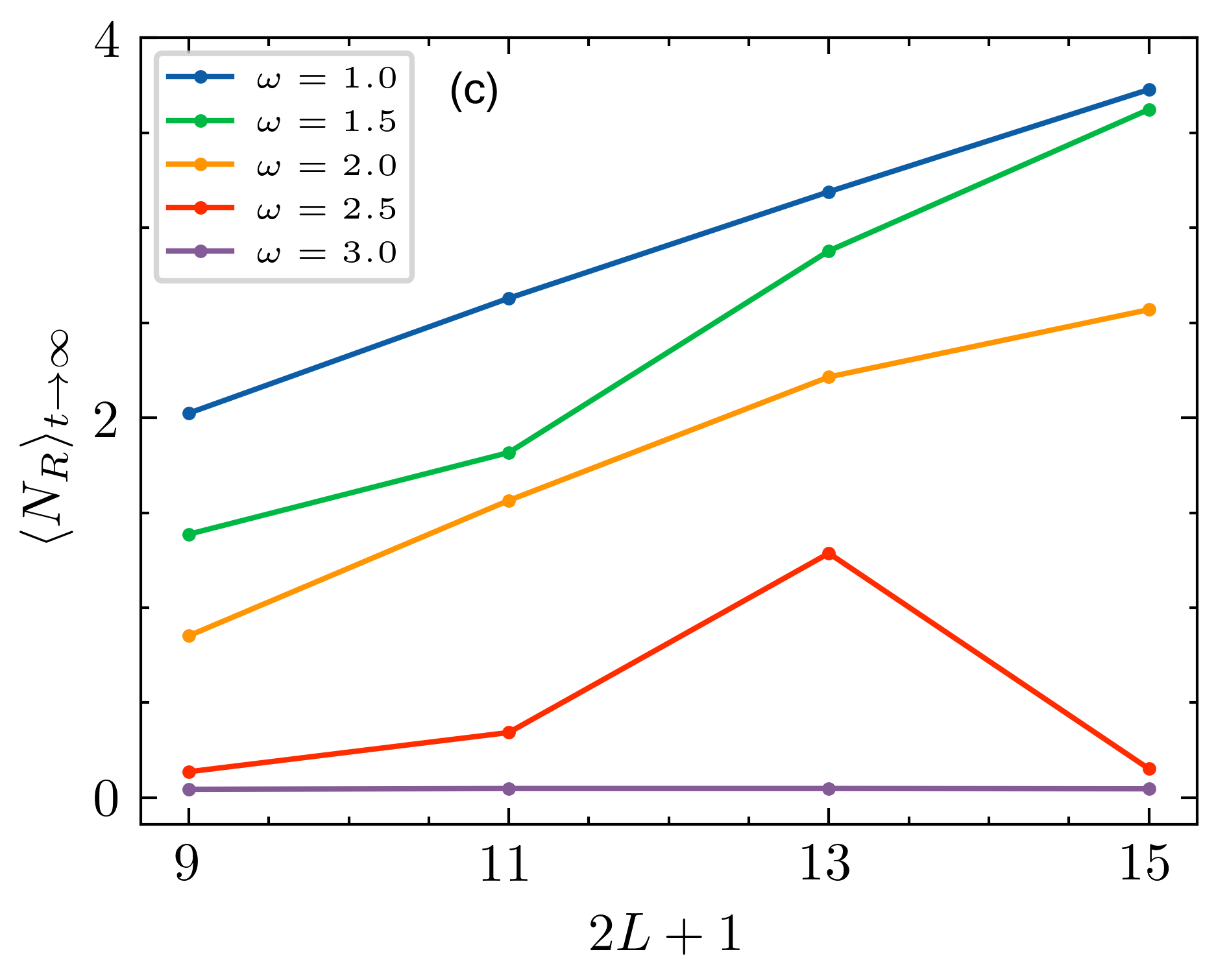}
    \includegraphics[width=0.49\textwidth]{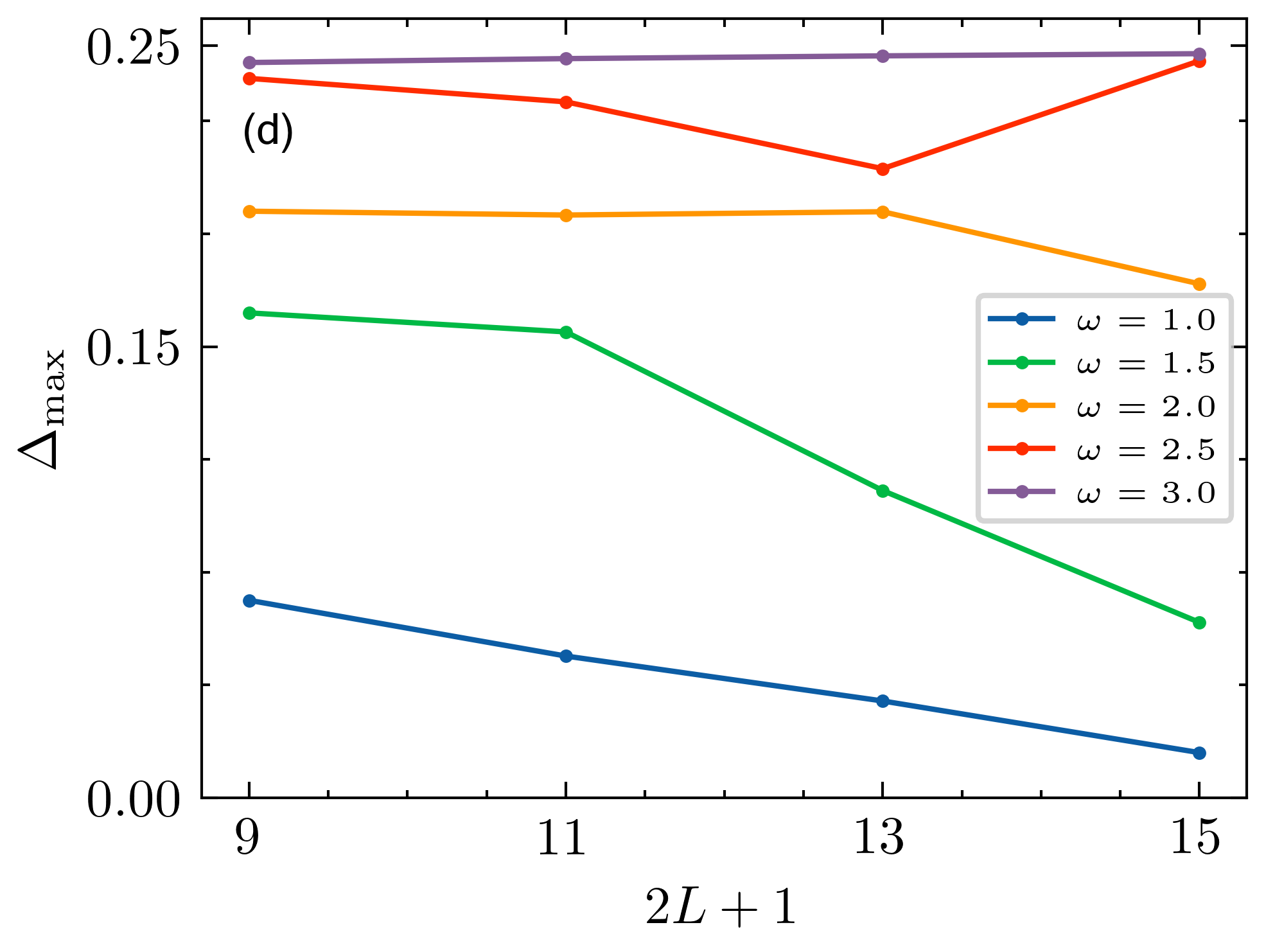}
    \caption{\label{fig:fourier}The same as in  Fig.~\ref{fig:alternating} except for the driving protocol $f_t = \frac{80}{\pi^2}\sum_{j=1}^{4}
    \frac{(-1)^{j+1}}{(2 j + 1)^2}
    \sin{ \omega t }$.}
\end{figure*}

Here we present results for various driving protocols $f_t$ in eq. \eqref{Vt}. The protocols under study are shown in Fig.~\ref{fig:ft}. The results are shown in  Figs.~\ref{fig:alternating}---~\ref{fig:fourier}. One can see that they are qualitatively similar to the results reported in the main text.

\section{S2. A QPC with on-site potential}

\begin{figure*}[ht]
    \includegraphics[width=0.48\textwidth]{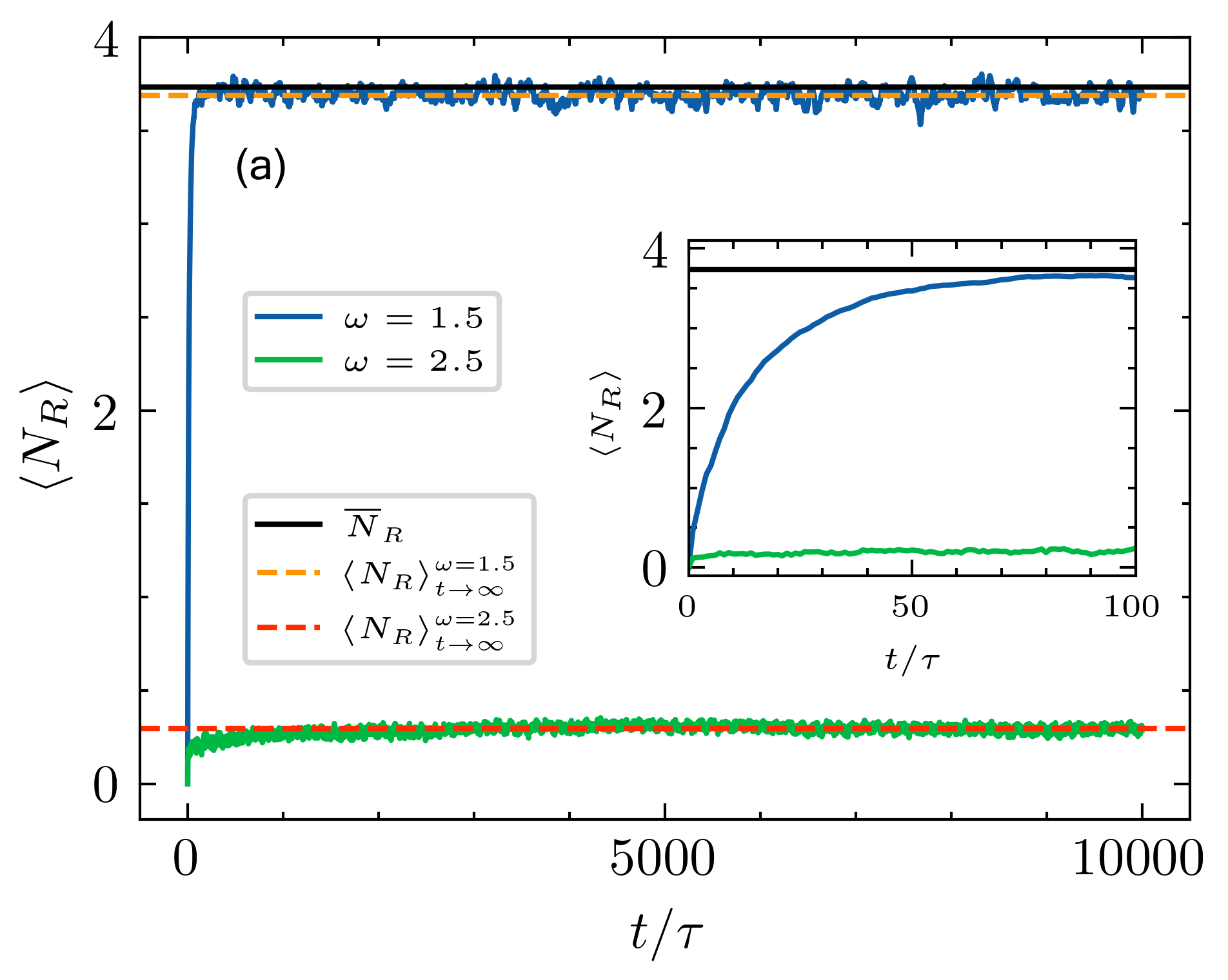}
    \includegraphics[width=0.49\textwidth]{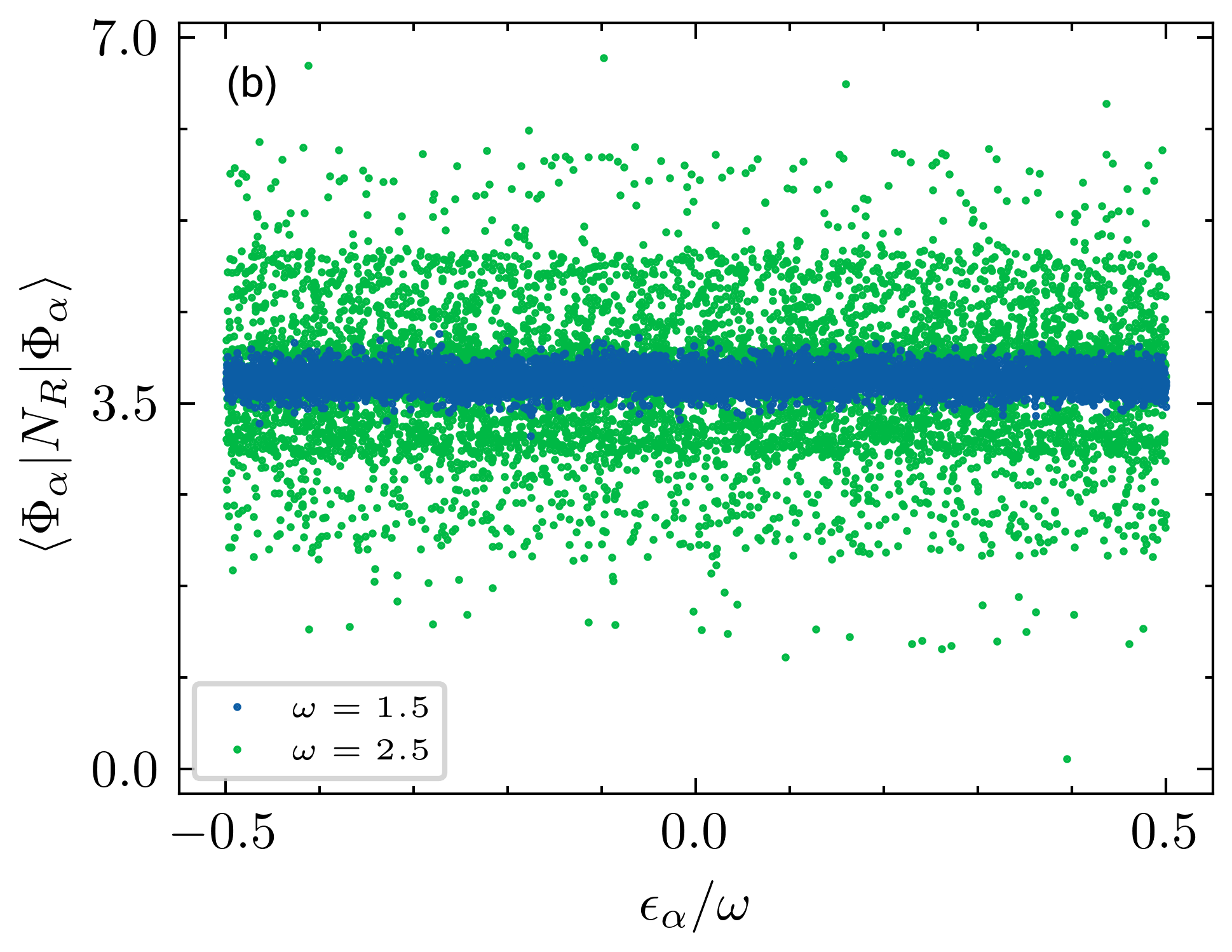}
    \\
\includegraphics[width=0.48\textwidth]{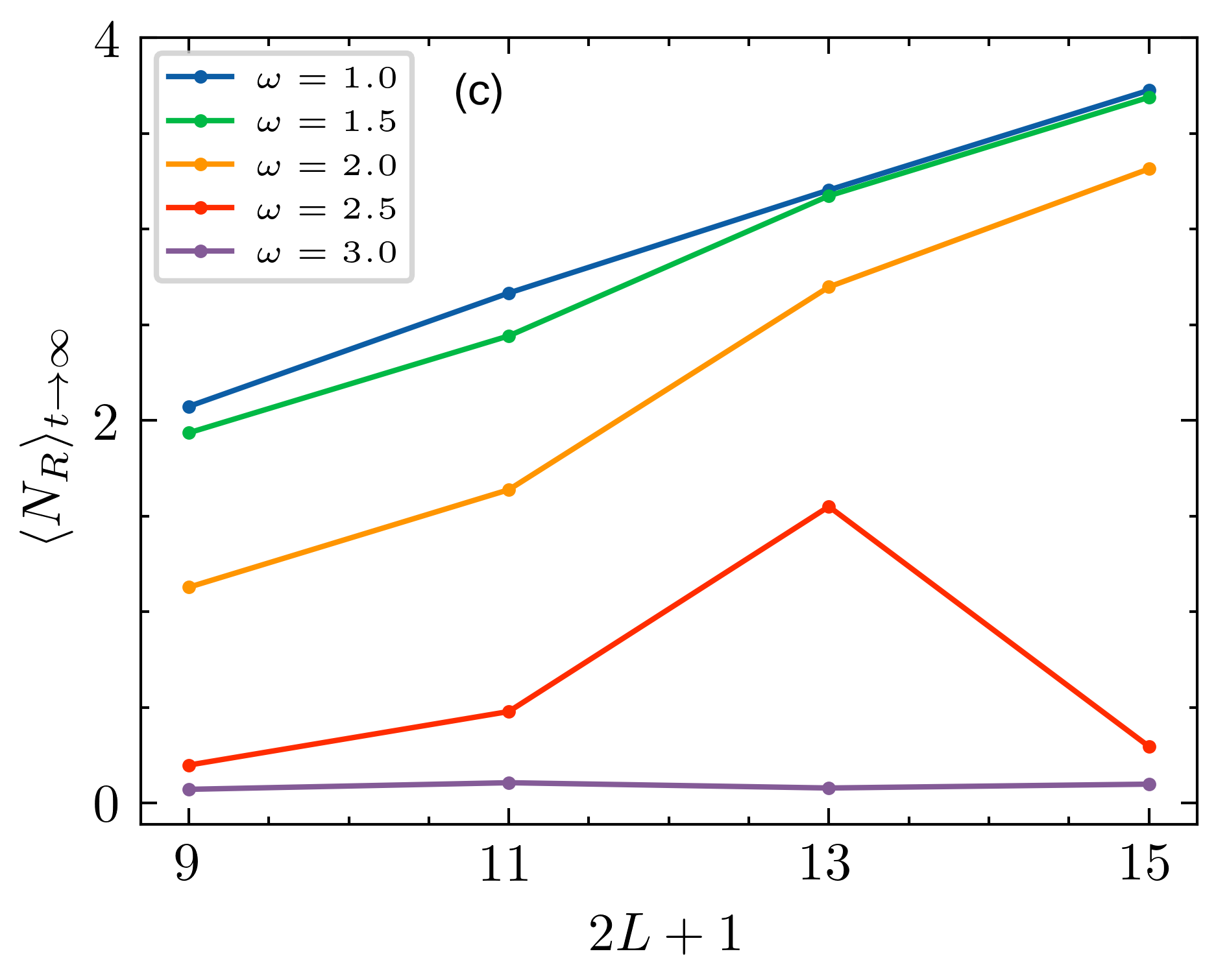}
    \includegraphics[width=0.49\textwidth]{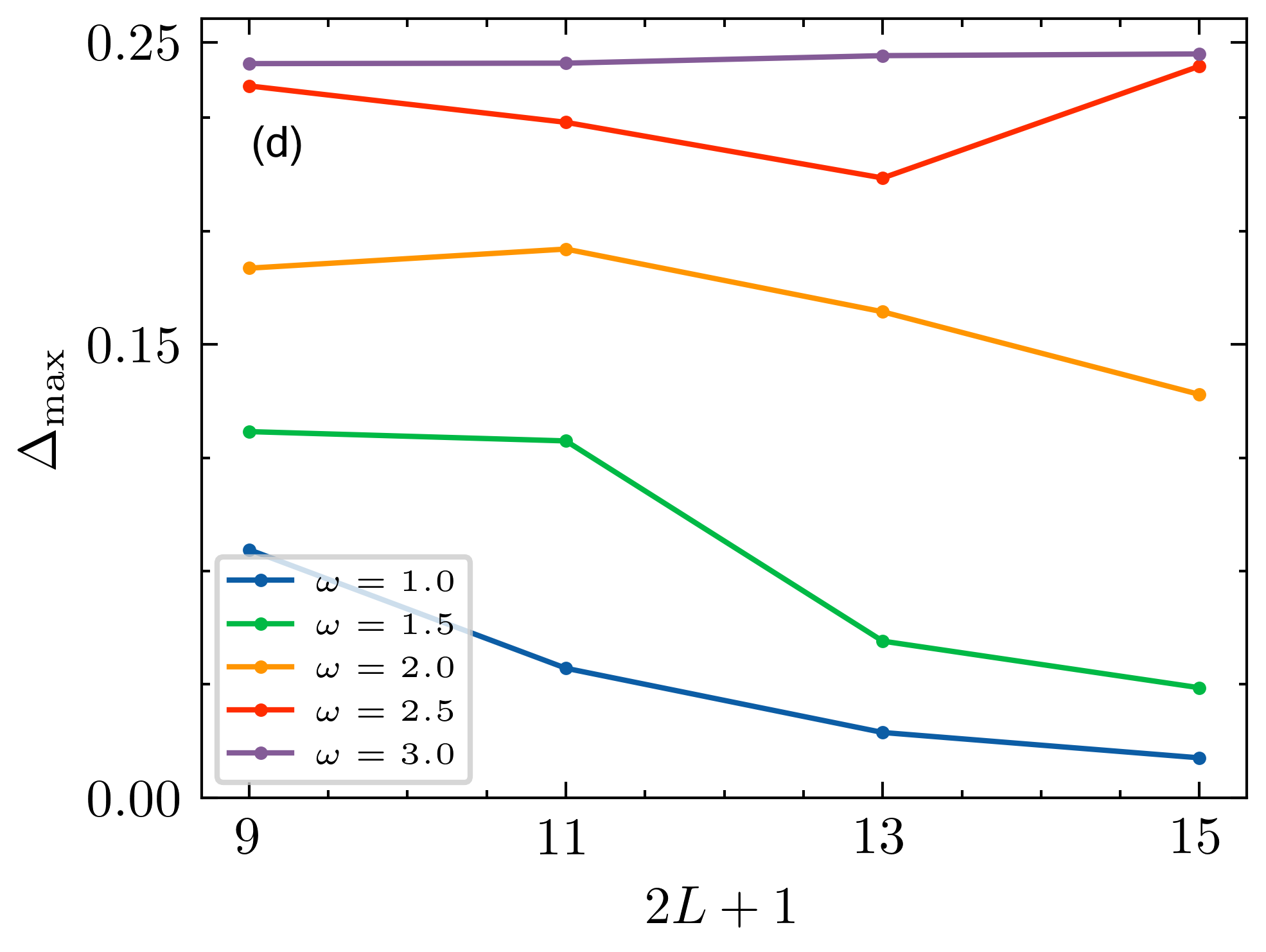}
\caption{\label{fig:conformal2}The same as in  Fig.~\ref{fig:alternating} except for the QPC Hamiltonian $V_t$ given by eq. \eqref{Vt on-site} with the tunneling amplitude $f_t = \sin{ \omega t }$ and on-site potential $g_t = \sin{ \omega t }$.}
\end{figure*}

It was found in ref. \cite{gamayun2021nonequilibrium} that, in the noninteracting case, an alternating on-site potential within the QPC typically destroys the effect and ensures the current flow at any frequency. However, a notable exception was found for the QPC term $V_t$ given by
\begin{equation}\label{Vt on-site}
    V_t = -\frac{1}{2}
    \begin{pmatrix}
  \hat{c}^\dagger_{L+1}& \hat{c}_{L+2}
 \end{pmatrix}
    \begin{pmatrix}
 g_t & f_t \\
 f_t & -g_t
 \end{pmatrix}
 \begin{pmatrix}
 \hat{c}^\dagger_{L+1} \\
 \hat{c}_{L+2}
 \end{pmatrix}.
\end{equation}
with the tunneling amplitude $f_t=\sin(\omega t)$ and on-site potential $g_t=\sin(\omega t)$. We have found the same qualitative picture in the interacting case. In particular, the current halts completely for the above $V_t$, as shown in Fig. ~\ref{fig:conformal2}.

\section{S3. Various interaction terms}
\begin{figure*}[ht]
  \includegraphics[width=0.48\textwidth]{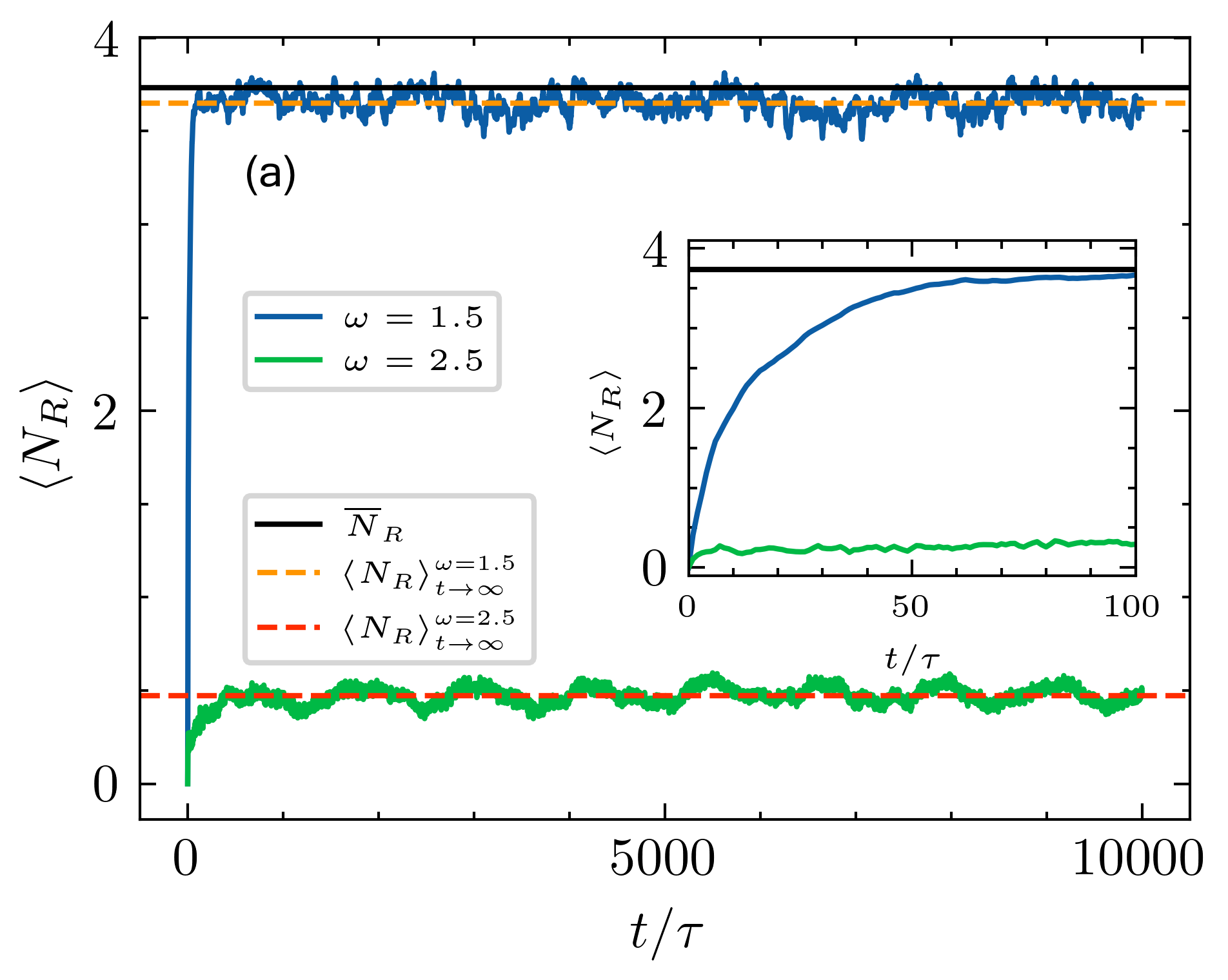}
    \includegraphics[width=0.49\textwidth]{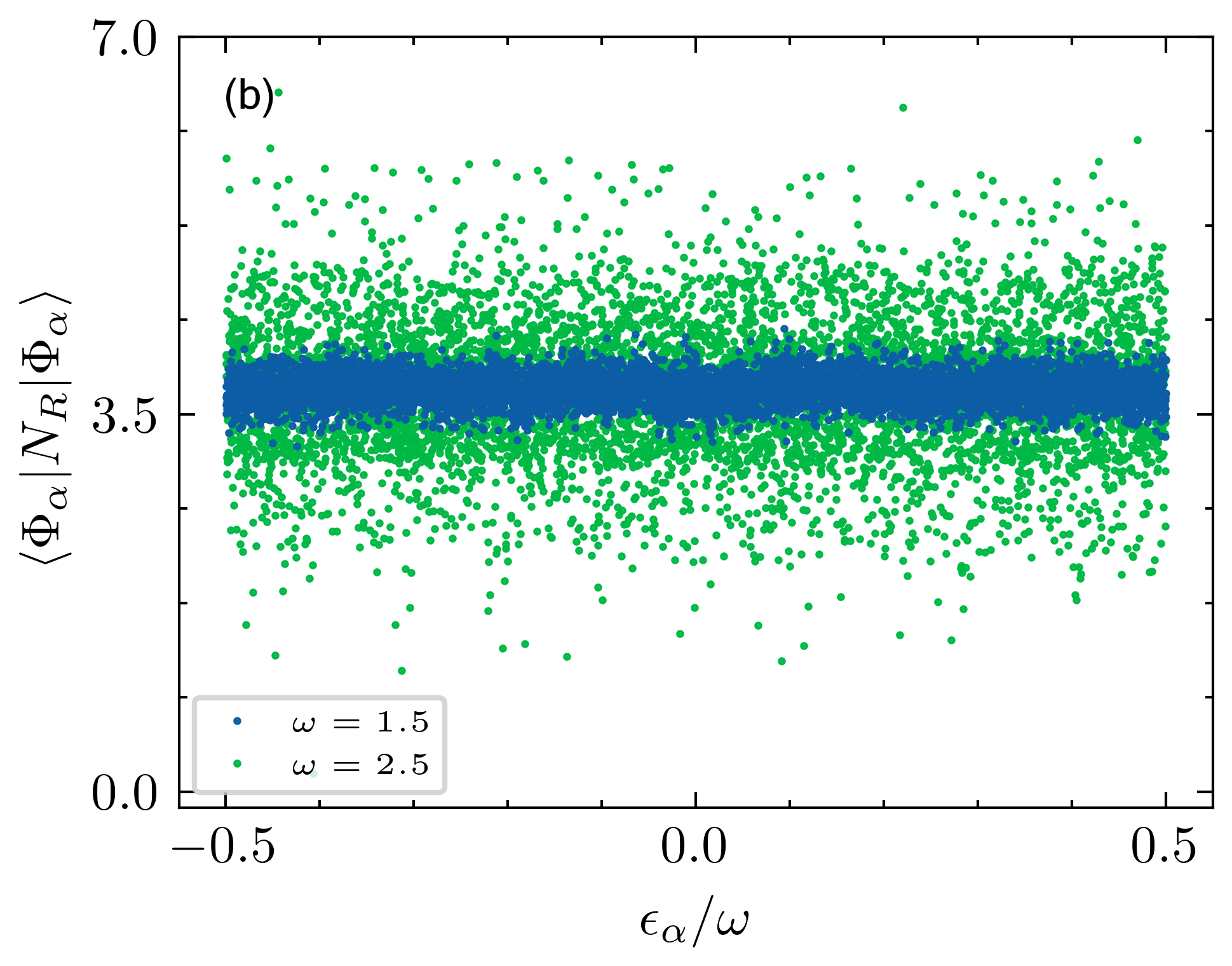}
    \\
     \includegraphics[width=0.48\textwidth]{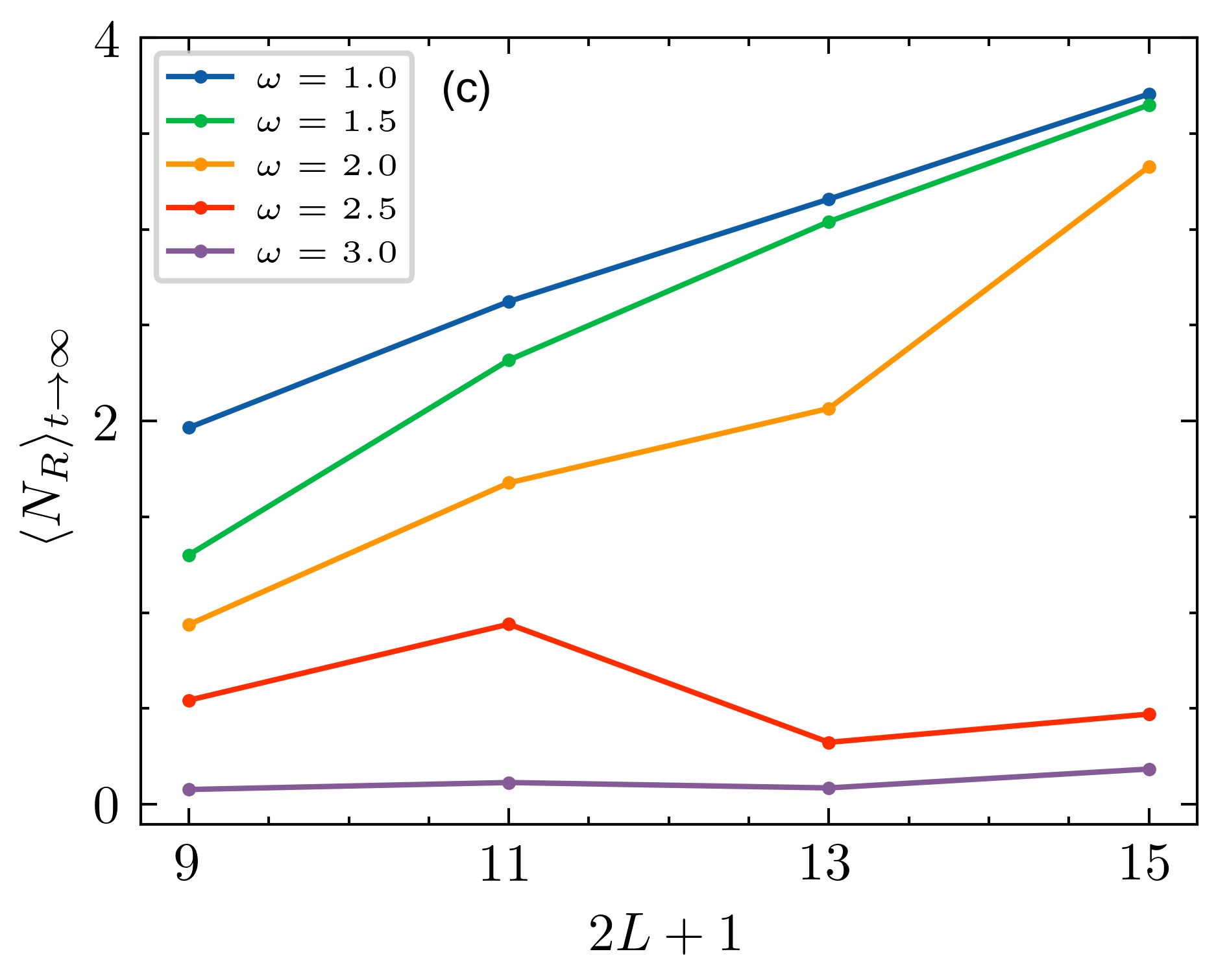}
    \includegraphics[width=0.49\textwidth]{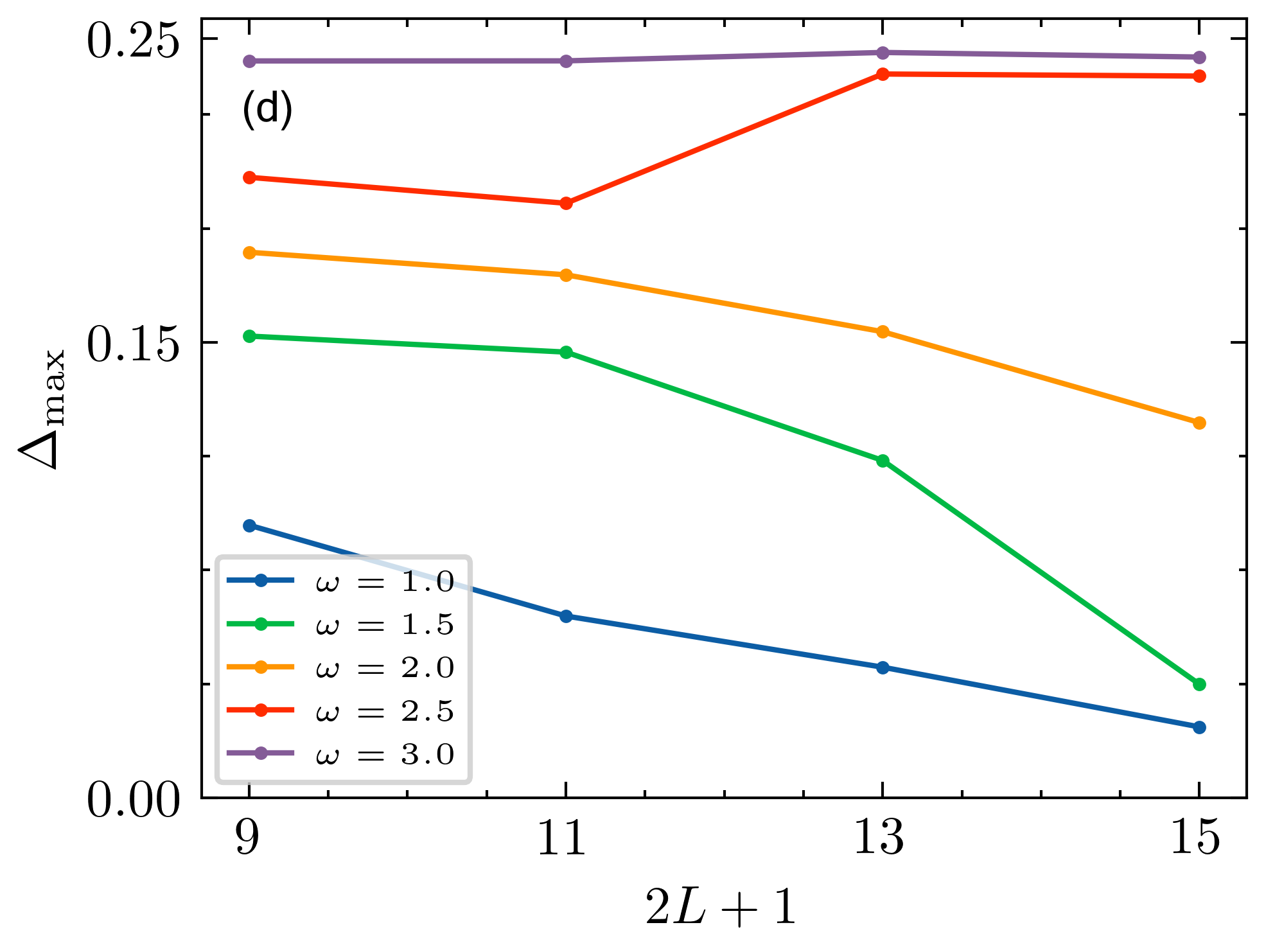}
\caption{\label{fig:next-nearest}The same as in  Fig.~\ref{fig:alternating} except for the driving protocol  $f_t = \sin\omega t$ and the interaction term $V_{\rm int}$ given by eq. \eqref{Vint NNN} with  $U=U'=0.5$.}.
\end{figure*}

We have verified that the physical picture remains the same for various types of interaction terms. As an illustration, we show in Fig.~\ref{fig:next-nearest} the results for
\begin{equation}\label{Vint NNN}
    V_{\text{int}}  =  U\sum_{j=1}^{2L}n_jn_{j+1}+U'\sum_{j=1}^{2L-1}n_jn_{j+2},
\end{equation}
where $U$ and $U'$ are two interaction constants.


\clearpage

\section{S4. Estimate of the prethermalization time and energy scales }

A conservative estimate $\Lambda$ of the prethermalization energy scale is obtained as the sum of operator norms of Hamiltonian terms with a support covering a fixed site  \cite{Ho_2023_Quantum}. For our system \eqref{H}-\eqref{Vint} this implies $\Lambda=1+2 U$.

The ratio of frequency-dependent thermalization timescales $\tau_{\rm th}(\omega)$ for two different frequencies is estimated as $\tau_{\rm th}(\omega_2)/\tau_{\rm th}(\omega_1)\sim \exp\left((\omega_2-\omega_1)/\Lambda\right)$. For $\omega_1=1.5$, $\omega_2=2.5$ and $U=0.5$ this estimate reads $\tau_{\rm th}(\omega_2)/\tau_{\rm th}(\omega_1)\sim \sqrt{e}\simeq1.6$. Thus the prethermalization timescales for these two frequencies should be  of the same order of magnitude. However, we clearly observe thermalization at the frequency $\omega_1$, with $\tau_{\rm th}(\omega_1)\sim 10$, while no  thermalization is observed at the frequency $\omega_2$ up to times $\sim 1000\, \tau_{\rm th}(\omega_1)$. This argument confirms  that the observed lack of thermalization can not be explained by ``mere'' prethermalization.

\end{document}